\documentclass[twocolumn,pra,amsmath,amssymb,superscriptaddress,longbibliography]{revtex4-1}

\usepackage{tikz}
\usepackage{pgfplots} 
\pgfplotsset{compat = newest} 
\usepgfplotslibrary{ternary} 
\usepackage[T1]{fontenc}
\usepackage{bbold}

\usepackage{enumitem}

\usepackage[citecolor=blue, urlcolor=blue, linkcolor=blue, colorlinks=true]{hyperref}

\newcommand{\subfigimg}[3][,]{%
\setbox1=\hbox{\includegraphics[#1]{#3}}
\leavevmode\rlap{\usebox1}
\rlap{\hspace*{-3pt}\raisebox{\dimexpr\ht1+.15\baselineskip}{#2}}
\phantom{\usebox1}
}

\makeatletter
\newcommand*\rel@kern[1]{\kern#1\dimexpr\macc@kerna}
\newcommand*\widebar[1]{%
  \begingroup
  \def\mathaccent##1##2{%
    \rel@kern{0.8}%
    \overline{\rel@kern{-0.8}\macc@nucleus\rel@kern{0.2}}%
    \rel@kern{-0.2}%
  }%
  \macc@depth\@ne
  \let\math@bgroup\@empty \let\math@egroup\macc@set@skewchar
  \mathsurround\z@ \frozen@everymath{\mathgroup\macc@group\relax}%
  \macc@set@skewchar\relax
  \let\mathaccentV\macc@nested@a
  \macc@nested@a\relax111{#1}%
  \endgroup
}
\makeatother

\usepackage{txfontsb}
\usepackage{times}
\usepackage{mathrsfs}
\usepackage{dsfont}

\definecolor{nice-blue}{HTML}{1F77B4}
\definecolor{nice-green}{HTML}{2CA02C}
\definecolor{nice-orange}{HTML}{FF7F0E}
\definecolor{nice-red}{HTML}{D62728}

\newcommand{\PRLsep}{\noindent\makebox[\linewidth]{\resizebox{0.5\linewidth}{1pt}{$\bullet$}}\bigskip}

\begin{document}
\title{Intrinsic sign problems in topological quantum field theories}
\author{Adam Smith}
\email{adam.smith@tum.de}
\affiliation{Department of Physics, TFK, Technische Universit{\"a}t M{\"u}nchen, James-Franck-Stra{\ss}e 1, D-85748 Garching, Germany}
\author{Omri Golan}
\affiliation{Department of Condensed Matter Physics, Weizmann Institute of Science, Rehovot 76100, Israel}
\author{Zohar Ringel}
\affiliation{Racah Institute of Physics,
The Hebrew University of Jerusalem, Israel}

\date{\today}

\begin{abstract}
   The sign problem is a widespread numerical hurdle preventing us from simulating the equilibrium behavior of various problems at the forefront of physics. Focusing on an important sub-class of such problems, bosonic $(2+1)$-dimensional topological quantum field theories, here we provide a simple criterion to diagnose intrinsic sign problems---that is, sign problems that are inherent to that phase of matter and cannot be removed by any local unitary transformation. Explicitly, \textit{if the exchange statistics of the anyonic excitations do not form complete sets of roots of unity, then the model has an intrinsic sign problem}. This establishes a concrete connection between the statistics of anyons, contained in the modular $S$ and $T$ matrices, and the presence of a sign problem in a microscopic Hamiltonian. Furthermore, it places constraints on the phases that can be realised by stoquastic Hamiltonians. We prove this and a more restrictive criterion for the large set of gapped bosonic models described by an abelian topological quantum field theory at low-energy, and offer evidence that it applies more generally with analogous results for non-abelian and chiral theories.
\end{abstract}

\maketitle


\section{Introduction}

Our theoretical and practical understanding of quantum systems involving many interacting particles often relies on our ability to simulate them efficiently. Indeed, many outstanding problems in physics such as High-Tc superconductivity~\cite{Bednorz1986,Lee2018a}, confinement transitions in quantum chromodynamics~\cite{Toussaint1990}, and topological quantum matter~\cite{Wen2017}, are those that are hard to simulate numerically. One of the most powerful set of tools for our understanding of equilibrium physics are Monte Carlo methods~\cite{Ulam1949,Assaad2008,Gubernatis2016}. However, despite their success for an increasingly large set of quantum systems, in many circumstances these methods are plagued by a numerical obstacle known as the sign problem~\cite{Loh1990}, which renders Quantum Monte Carlo (QMC) algorithms intractable. 

From a practical standpoint a model with a sign problem is one for which, after considerable effort, no representation of the partition function has been found such that the model appears as a proper statistical mechanical model with non-negative real Boltzmann weights~\cite{Gubernatis2016,Assaad2008}. Having a proper Statistical Mechanical representation is desirable since for such models the distribution of observables can be efficiently sampled from in polynomial time using Markov Chain Monte Carlo~\cite{Hastings1970}. In the field of quantum computation, a well known class of bosonic Hamiltonians without a sign problem are called stoquastic Hamiltonians~\cite{Bravyi2008}. In accordance with the fact that they can be simulated efficiently using QMC, adiabatic computation using stoquastic Hamiltonians is believed to give rise to a weaker computational complexity class (postBPP) compared to generic Hamiltonians that have a sign problem~\cite{Bravyi2008}.

Research focusing on finding solutions to sign problems has a long and successful history. However, the complementary question, of whether there are fundamental obstructions to solving the sign problem for certain phases of matter, began receiving attention only recently. It has been argued that a generic solution to the sign problem is unlikely from complexity theory perspective~\cite{Iazzi2016,Troyer2005,Marvian2018}. Furthermore, Hastings~\cite{Hastings2016} has proven that a specific lattice gauge theory (the doubled semion model) does not admit a solution for the sign problem, conditioned that the Hamiltonian is made of commuting projectors. In addition it was shown by Ringel and Kovrizhin~\cite{Ringel2017} that bosonic chiral topological phases with a quantized thermal Hall conductance---or equivalently a gravitational anomaly---have a sign problem. The latter work can be seen as an example of an intrinsic sign problem: a sign problem that is an inherent property of the phase of matter, not conditioned on microscopic constraints.

An additional motivation to reveal intrinsic sign problems comes from the recent interest in quantum supremacy~\cite{Boixo2018,Arute2019}. Indeed having a quantum device that can efficiently and accurately simulate a model with an intrinsic sign problem can be considered as evidence for a practical computation advantage. This may also provide a deeper understanding of quantum and classical complexity classes~\cite{Nielsen2010}.

The current work focuses on intrinsic sign problems in the context of bosonic $(2+1)$-dimensional topological quantum field theories~\cite{Atiyah1988,Witten1989,SteveTopo}. Topological quantum field theory is a powerful analytical framework for describing various exotic phases of matter with long-range entanglement. These arise in the context of fractional quantum Hall physics~\cite{Simon1998,Hansson2017}, quantum spin-liquids~\cite{Savary2017,Knolle2019}, and lattice gauge theories~\cite{Oeckl2005}. Some of these phases, if realized, may serve as platforms for quantum computers via the method of topological quantum computation using the adiabatic braiding of non-abelian anyons~\cite{Sarma2006,Nayak2008,Lahtinen2017}. As our ability to design and control materials exhibiting these phases relies heavily on simulations, it is desirable to understand which TQFTs have intrinsic sign problems. This is especially relevant in light of the recent experimental demonstration of fractional anyon statistics~\cite{Bartolomei2020}. However, apart from Hastings' work~\cite{Hastings2016}, it remains unclear whether TQFTs more generally lead to intrinsic sign problems and whether this can be diagnosed based on the TQFT data: the $T$ and $S$ matrices defining the exchange and mutual statistics of the anyonic quasi-particles in the theory. 

A closely related question that has recently received significant interest is how to extract topological data, such as the above $S$ and $T$ matrices, from ground state wavefunctions~\cite{Zhang2012,Cincio2013,Moradi2014,Mei2014,Zhu2018,Zhu2017a}. At the core of these works is the connection between the statistics of excitations and the modular transformations of the torus generated by Dehn twists~\cite{CFTBook}. For example, one approach is to implement these Dehn twists by reconnecting the lattice either adiabatically~\cite{You2015}, or instantaneously~\cite{Zhou2018}. Another is to compute the inner product between rotated or sheared minimum entropy states (MES)~\cite{Zhang2012,Zhu2013}. We also note the work of Haah~\cite{Haah2016}, where the $S$ matrix is extracted from a twisted product of ground state density matrices, which was a central element in the sign problem proof by Hastings~\cite{Hastings2016}. In this paper we develop new geometrical and analytical tools to extract topological information from ground states. These tools form a central part of the proof of our main result.

In this work we establish that some topologically ordered models indeed have intrinsic sign problem and point to its physical source: the statistics of the quasi-particles. 
Specifically, we will prove the following result:
\begin{center}
\parbox{.9\columnwidth}{
\emph{Let $\hat{H}$ be a stoquastic gapped non-chiral bosonic Hamiltonian in two dimensions with an abelian TQFT description at low energy, then the topological spins form complete sets of roots of unity.}
}
\end{center}
This provides a simple criterion for diagnosing intrinsic sign problems in topological models. As a corollary we also have the more general criterion that \emph{there exists a ground state basis for $\hat{H}$ with respect to which all modular transformations are non-negative.} This means, that for the Hamiltonian to be stoquastic, the modular $S$ and $T$ matrices that define the TQFT can be made simultaneously non-negative. While we establish these criteria for non-chiral and abelian models, we also obtain analogous results for chiral and non-abelian phases. In a parallel work we also establish a variant of our results that applies for fermionic Hamiltonians~\cite{Golan2020}. 

Our results extend far beyond previously established results on intrinsic sign problems in two key aspects: they apply to a much larger set of TQFTs, and they apply beyond commuting projector Hamiltonians, thereby establishing a direct relation between physical properties of the phase and the sign problem. These results additionally place fundamental constraints on the phases that can be realised by stoquastic Hamiltonians. Importantly, however, our results do preclude solutions to central open problems in many-body physics such as high-$T_c$ superconductivity, or quantum spin liquids in frustrated magnets. Additionally, it is believed that a wide class of symmetry protected topological phases should be free from sign problems~\cite{Geraedts2013,Bondesan2017}.

\section{Examples of topological intrinsic sign problems}\label{sec: examples}

Let us briefly consider two examples to demonstrate the above results. The first is the toric code, where there is no sign-problem, and the second is the double semion model, where it has already been proven that there is an intrinsic sign problem~\cite{Hastings2016}.

The toric code has the $S$ and $T$ matrices
\begin{equation}
    S_\text{TC} = \frac{1}{2} \left(
    \begin{array}{cccc}
    1 & 1 & 1 & 1 \\
    1 & 1 & -1 & -1 \\
    1 & -1 & 1 & -1 \\
    1 & -1 & -1 & 1
    \end{array}\right),
    \qquad T_\text{TC} = \left(
    \begin{array}{cccc}
    1 & 0 & 0 & 0 \\
    0 & 1 & 0 & 0 \\
    0 & 0 & 1 & 0 \\
    0 & 0 & 0 & -1
    \end{array}\right).
\end{equation}
The topological spins $\{\theta_a\}_{a \in \mathbb{A}} = \{1,1,1,-1\} = \{1\} \cup \{1\} \cup \{1,-1\}$ do come in complete sets of roots of unity and so the toric code fulfills our criteria. Although this is not a sufficient condition, the toric code is indeed stoquastic in the standard spin (qubit basis). The $S$ and $T$ matrices can be made simultaneously non-negative with the unitary $V = \mathds{1}_2\otimes H$, where $H$ is the $2\times 2$ Hadamard matrix, resulting in the matrices
\begin{equation}
    VS_\text{TC}V^\dag = \frac{1}{2} \left(
    \begin{array}{cccc}
    1 & 0 & 0 & 0 \\
    0 & 0 & 1 & 0 \\
    0 & 1 & 0 & 0 \\
    0 & 0 & 0 & 1
    \end{array}\right),
    \quad VT_\text{TC}V^\dag = \left(
    \begin{array}{cccc}
    1 & 0 & 0 & 0 \\
    0 & 1 & 0 & 0 \\
    0 & 0 & 0 & 1 \\
    0 & 0 & 1 & 0
    \end{array}\right).
\end{equation}

The double semion model on the other hand has the $S$ and $T$ matrices
\begin{equation}
    S_\text{DS} = \frac{1}{2} \left(
    \begin{array}{cccc}
    1 & 1 & 1 & 1 \\
    1 & -1 & 1 & -1 \\
    1 & 1 & -1 & -1 \\
    1 & -1 & -1 & 1
    \end{array}\right),
    \qquad T_\text{DS} = \left(
    \begin{array}{cccc}
    1 & 0 & 0 & 0 \\
    0 & i & 0 & 0 \\
    0 & 0 & -i & 0 \\
    0 & 0 & 0 & 1
    \end{array}\right).
\end{equation}
Here the topological spins $\{\theta_a\}_{a \in \mathbb{A}} = \{1,i,-i,1\}$ do not form complete sets of roots unity. Therefore the double semion model fails our criteria and has an intrinsic sign problem, consistent with Ref.~\cite{Hastings2016}.

The toric code and double semion model are the two possible abelian string-net models~\cite{Freedman2004,Levin2005} built on a $\mathbb{Z}_2$ input theory and out criteria apply much more generally. It is also possible to list all the abelian $\mathbb{Z}_N$ string-net models and extract their $S$ and $T$ matrices, see Ref.~\cite{Lin2014}. For a $\mathbb{Z}_N$ input theory there are $N$ possible string-net models, which are labelled by $\mathbb{Z}^p_N$, where $p = 0,\ldots,N-1$. The models with $p=0$ correspond to higher-order generalizations of the toric code and are trivially stoquastic. However, in the appendix we prove that for all $N\geq2$ and $p\neq 0$, the $\mathbb{Z}^p_N$ string-net models have intrinsic sign problems. We also note that, for all of these models the intrinsic sign problem can be diagnosed from the spectrum of the $T$-matrix alone. This amounts to a complete classification of intrinsic sign problems in $\mathbb{Z}_N$ string-net models.

This is not an exhaustive list of abelian TQFTS, which include for instance the $\mathbb{Z}_{N_1}\times \cdots \times \mathbb{Z}_{N_m}$ abelian string-net models, whose topological spins can be found in Ref.~\cite{Lin2014} and easily checked. We also provide examples for chiral and non-abelian cases in Sec.~\ref{sec: chrial}.

\section{Outline of this paper}\label{sec: sketch}

Before delving into details of a proof, we would first like to outline our arguments. The main goal of the paper is to establish a connection between intrinsic sign problems for microscopic Hamiltonians---that is, sign problems that can't be removed---and of the properties of the topological phase of matter that they realise. Our proof can be viewed in two ways. First, that if we have a microscopic Hamiltonian in two dimensions that does not have an intrinsic sign problem, then there are restrictions on the topological phases it can realise. Second, if we have a Hamiltonian that realises a topological phase that fails our criteria, then the Hamiltonian has an intrinsic sign problem.

The modular transformations are important properties of a TQFT. These transformations are generated by the modular matrices $S$ and $T$, which define a topological quantum field theory (TQFT) and correspond to the modular transformations shown in Fig.~\ref{fig: modular transformations}(b-c). The matrix elements $S_{ab}$ contain the mutual statistics of anyons of type $a$ and $b$, and $T_{ab} = \theta_a\delta_{ab}$ are the topological spins corresponding to the exchange statistics of type $a$. In this paper we prove that the eigenvalues of the $T$-matrix must come in complete sets of roots of unity if the Hamiltonian is free from an intrinsic sign problem. Our arguments can then also be repeated for the matrix $T' = STS^{-1}$ (more commonly denoted $U$~\cite{CFTBook}, but we reserve this letter for elsewhere). The two matrices, $T$ and $T'$, contain the same information as $S$ and $T$ and generate all modular transformations. We introduce the most important properties of TQFTs and of topologically ordered Hamiltonians in Sec.~\ref{sec: setup}, with more details included in Appendix.~\ref{ap: TQFT}.

In Sec.~\ref{sec: Dehn twists} we introduce a microscopic prescription for extracting the $T$-matrix from the ground states of a Hamiltonian defined on a torus. The modular $T$-matrix corresponds to a Dehn twist on the torus. A Dehn twist is performed by cutting open the torus and twisting by a full turn before gluing back together, see Fig.~\ref{fig: modular transformations}(b). While this prescription for the Dehn twist is well defined for a continuum model, its definition is more subtle in the context of the lattice. Our microscopic prescription consists of two parts $\hat{T} = \hat{U}\hat{T}_g$. The first part $\hat{T}_g$ corresponds to the naive geometric implementation of the Dehn twist, by cutting and twisting the lattice. The second part $\hat{U}$ is included to fix the lattice distortions that are introduced by $\hat{T}_g$, as shown in Fig.~\ref{fig: DehnTwist lattice}. A similar lattice fixing procedure was considered in Ref.~\cite{Zhu2018} for string-net models, however, we need a more general procedure and one where we can keep track of signs induced by $\hat{U}$. Importantly, our implementation of $\hat{U}$ ensures that this lattice fixing is done locally, adiabatically and in a sign-free manner. It therefore only modifies microscopic details of the ground states, but does not change the topological, long-range properties.

The bulk of the paper, in Sec.~\ref{sec: LSAP}, is then devoted to defining the operator operator $\hat{U}$ and proving that on the level of the TQFT it acts as the identity, and only modifies microscopic details. The starting point is a skewed lattice connectivity left behind by $\hat{T}_g$. The lattice fixing implemented by $\hat{U}$ then proceeds by the following steps: (i) create two nearby lattice disclocations; (ii) move these dislocations around the torus leaving behind a string of reconnected bonds of the lattice; (iii) remove the dislocations when they meet on the far side of the torus; (iv) repeat the steps (i-iii) for the neighbouring loops until the entire lattice is reconnected correctly. Each step is implemented adiabatically and by local perturbations such that we preserve stoquasticity at every step. In the simplest case, this procedure fixes the lattice exactly as intended. The situation is complicated slightly if the dislocations bind a topological flux/excitation but we are able to simply modify our procedure to account for this possibility.

In Sec.~\ref{sec: T non-negative} we combine the properties of $\hat{T}_g$ and $\hat{U}$ to show that if the Hamiltonian does not have an intrinsic sign problem, then there exists a ground state basis $\{|\alpha\rangle \}$ such that $\langle\beta | \hat{T} |\alpha \rangle \geq 0$. Since $\hat{T}$ is unitary in the ground state sub-space, this implies that it is a permutation matrix in this basis and so its eigenvalues form complete sets of roots of unity, see Appendix.~\ref{ap: permutations}. These eigenvalues are precisely the topological spins $\theta_a$ of the anyons in the model, and so we must have $\{\theta_a\}_{a \in \mathbb{A}} = S_{m_1} \cup \cdots \cup S_{m_k}$, where $S_{m} = \{e^{2\pi i j/m} \;|\; j = 0,\ldots,m-1\}$. By extension we also show that any modular transformation, including $T$ and $S$, are non-negative in this basis. Therefore, we have the following more restrictive criteria: if there does not exist a basis such that $S$ and $T$ are non-negative, then the Hamiltonian realising this phase has an incurable sign problem.

In Secs.~\ref{sec: setup}--\ref{sec: T non-negative} we treat only non-chiral models that host abelian anyons. This is in order to make our arguments more precise and to avoid the additional subtleties introduced by chiral and non-abelian models. In Sec.~\ref{sec: chrial} we show that analogous results also hold for the more general case of chiral and non-abelian models, and also consider finite-size effects. To establish these more general results we use additional more physical arguments and numerical evidence from the literature. Finally we close with a discussion and outlook in Sec.~\ref{sec: discussion}.

\section{Setup}\label{sec: setup}

To set up a proof of our result we first make some definitions, namely what it means to have an intrinsic sign problem, the type of Hamiltonian that we consider, our basis choices, and a prescription for a Dehn twist on a lattice. In this paper we consider two-dimensional gapped bosonic Hamiltonians that are described by a topological quantum field theory (TQFT) at low energy. We will study these Hamiltonians on a torus where we have a degenerate ground state manifold. The non-trivial topology of the lattice is used as a tool allowing us to extract the topological information about the model but the intrinsic sign problem we are concerned with persists in any geometry.

\subsection{Intrinsic sign problems}

A Hamiltonian is \emph{stoquastic} with respect to a given basis if all of its off-diagonal elements are non-positive~\cite{Bravyi2008}. We will further restrict to Hamiltonians that are both local and locally stoquastic---i.e. Hamiltonians of the form $\hat{H} = \sum_i \hat{h}_i$, where $\hat{h}_i$ is non-trivial on a region with finite support with radius at most $R_H$, and each $\hat{h}_i$ is itself stoquastic. If the Hamiltonian is not stoquastic then it has a \emph{sign problem}.
We say that the sign problem can be removed if there exists a local unitary transformation (finite-depth and finite-range quantum circuit) to a new basis such that the Hamiltonian is locally stoquastic. We call this the \emph{computational basis} and label the basis states by Roman letters, e.g., $|i\rangle, |j\rangle$ etc. In this paper we are concerned with the case where such a computational basis does not exist and we have an \emph{intrinsic sign problem}~\cite{Hastings2016}, i.e., a sign-problem that cannot be removed by any local unitary transformation.

\subsection{TQFT Hamiltonians}\label{sec: TQFT H}

There have been a several works providing constructive definitions of two-dimensional Hamiltonians that have low energy TQFT descriptions, most famously the Turaev-Viro models~\cite{Turaev1992}, Kitaev's quantum double~\cite{Kitaev2003}, and the string-net constructions~\cite{Freedman2004,Levin2005}. In order to cover a larger class of models and provide model agnostic procedures, we provide here a general non-constructive definition of what it means for a microscopic to have a low energy TQFT description. This definition follows closely Ref.~\cite{Beverland2016}. We cover here the essential details to make this definition, but discuss TQFTs in more detail in Appendix.~\ref{ap: TQFT}.

A $(2+1)$-dimensional TQFT assigns a vectors space to every two-dimensional surface of a three-dimensional manifold (and assigns a map to every cobordism)~\cite{Atiyah1988,SteveTopo}. This vector space corresponds to the ground state sub-space of our Hamiltonian. These ground states---and the elementary excitations of the model---can be labelled by a set of anyon labels $\mathbb{A} = \{1,a,b,\ldots\}$. The content of the TQFT is defined by the modular matrices, $S$ and $T$, which correspond to modular transformations of the surface (see Sec.~\ref{sec: Dehn twists}) and contain the mutual and exchange statistics of the anyons, respectively. We consider the case where $S$ and $T$ are unitary and the input of the TQFT defines a unitary modular tensor category (UMTC). We then define \emph{a Hamiltonian to have a low energy TQFT description if there exists a set of Wilson loop operators $\{\hat{W}_a(C)\}_{a\in \mathbb{A}}$ for each directed closed curve $C$ that form a faithful representation of the Verlinde algebra}~\cite{Verlinde1988}, see Appendix~\ref{ap: TQFT} for more details. For the majority of this paper we consider the Wilson operators to act non-trivially on a finite neighbourhood, $R_W$, of the curve $C$ and commute exactly with the Hamiltonian. We will relax these properties later in Sec.~\ref{sec: non-commuting}, where we consider Wilson operators with exponentially decaying tails, and inexact commutation with the Hamiltonian on a finite system. The Wilson loop operators have the interpretation of creating an anyon-antianyon pair and dragging the anyon around the closed loop $C$ before re-annihilating the pair.

\begin{figure}[t]
    \centering
    \includegraphics[width = \columnwidth]{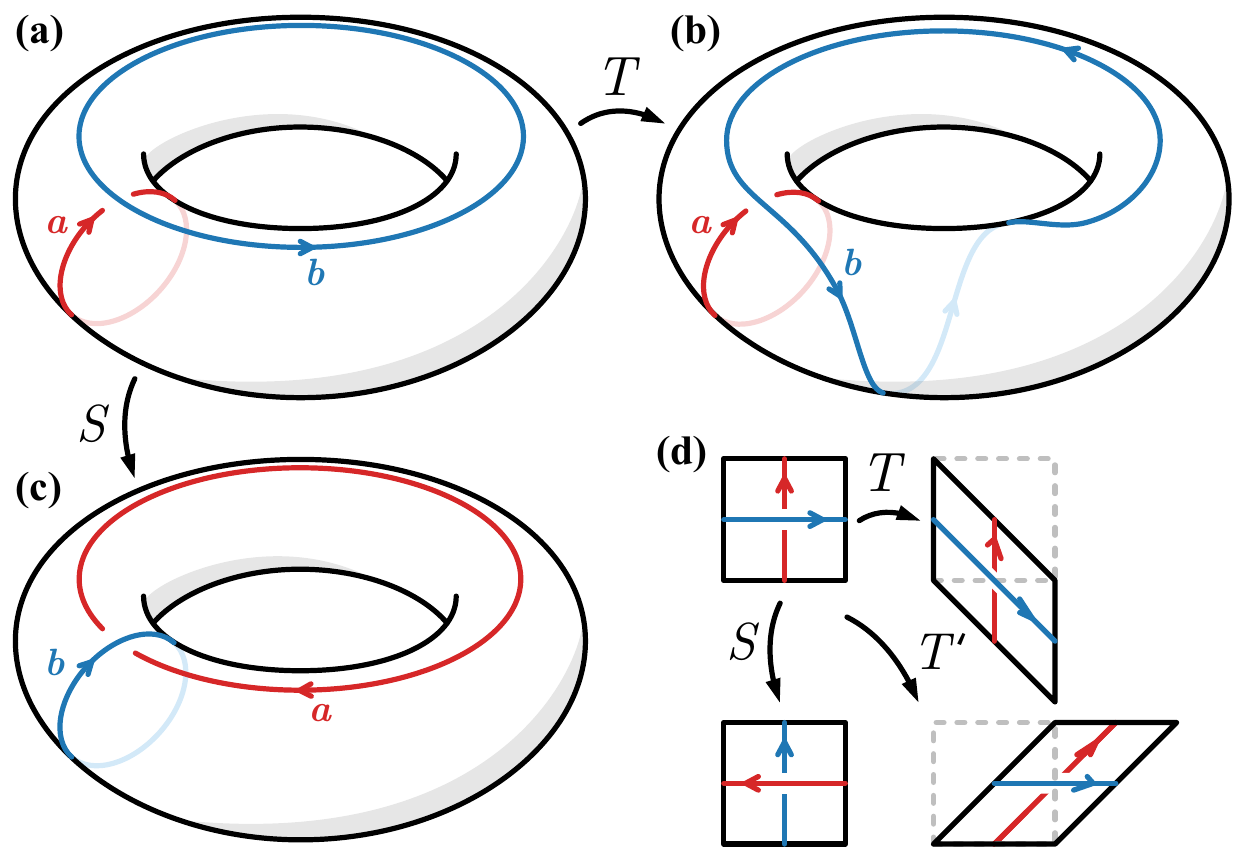}
    \caption{Modular transformations for the Torus shown in \textbf{(a)} generated by: \textbf{(b)} the $T$-matrix and \textbf{(c)} the $S$-matrix. The action of the modular transformations can defined in terms of the expectation values of Wilson loop operators around the two handles. \textbf{(d)} Modular transformations $S, T, T'$ for the square representation of the torus with identified edges.}
    \label{fig: modular transformations}
\end{figure}

\subsection{Basis choices}

From now on we assume that we have a locally stoquastic Hamiltonian $\hat{H}$ that has a low energy TQFT description, as defined in the previous two sections. Turning to basis choices for the ground state sub-space for the torus, the TQFT assumptions and stoquasticity for our Hamiltonian allow us to make two particular choices: the \emph{canonical basis} and the \emph{ergodic basis}.

The \emph{canonical basis}~\cite{Dittrich2017,Aasen2017,Koenig2010}, we will label by Roman letters at the start of the alphabet, e.g., $|a\rangle, |b\rangle$. This actually refers to two choices of basis related by the $S$-matrix, depending on which non-contractible direction of the torus we define them with respect to, corresponding to the so-called Minimal Entropy States (MES)~\cite{Zhang2012}. Without loss of generality we will use the vertical canonical basis in the following. We define the basis in terms of the Wilson loops wrapping around the horizontal (longitudinal) and vertical (meridian) non-contractible loops of the torus. We start by fixing the first element $|a=1\rangle$ such that it is a simultaneous eigenvector of all the vertical Wilson loops with eigenvalue given by the quantum dimension $d_a$ (which equals 1 for abelian theories), i.e., $\hat{W}^v_a |1\rangle = d_a|1\rangle$. We can then generate the other ground states using the horizontal Wilson loops, $|a\rangle = \hat{W}^h_a |1\rangle$. In the Appendix~\ref{ap: orthogonality} we show that these form an orthonormal basis for the ground state sub-space. The state $|a\rangle$ can be viewed as having flux of type $a$ threading the vertical (meridian) loop. Alternatively, if we cut the torus open along the vertical loop we would get an excitation of type $a$ on one edge and $\bar{a}$ on the opposite edge. Equivalently, the state $a$ is the eigenstate of the Kirby loop projector $\hat{\Omega}^v_a = d_a\mathcal{D}^{-1}\sum_b S^*_{ab} \hat{W}^v_b$, i.e., $\hat{\Omega}^v_a|b\rangle = \delta_{ab} |b\rangle$~\cite{Beverland2016}.

To define the \emph{ergodic basis} we start by noting that since our Hamiltonian is stoquastic, the matrix $[-\hat{H} + \Lambda \hat{\mathds{1}}]_{ij}$ is element-wise non-negative if we choose $\Lambda>0$ sufficiently large. The term $\Lambda \hat{\mathds{1}}$ only shifts the spectrum and so the ground states of $\hat{H}$ are the same as the eigenstates of $-\hat{H} + \Lambda \hat{\mathds{1}}$ with largest eigenvalue. By the Frobenius-Perron theorem we can choose a set of orthogonal eigenvectors that are element-wise non-negative spanning the ground state sub-space, see Appendix~\ref{ap: Frobenius-Peron}.
These states, labelled by Greek letters, e.g., $|\alpha\rangle,|\beta\rangle$, correspond to different ergodic sectors---they have distinct support in the computational basis, i.e., $\langle\alpha|i\rangle\langle i | \beta\rangle = 0$ for all $i$ and $\alpha \neq \beta$. Given the canonical basis states, the ergodic states are specified by a unitary matrix, $V_{\alpha b}$, which allows us to label the ergodic states, i.e., $| \alpha\rangle = \sum_b V_{\alpha b} |b\rangle$.

In summary, we will use the following short hand for three different bases,
\begin{equation}
    |i\rangle \equiv |i\rangle_{\text{comp}}, \quad |a\rangle \equiv |a\rangle_{\text{can}}, \quad |\alpha\rangle \equiv |\alpha\rangle_{\text{erg}},
\end{equation}
for the computational $(i,j,\dots)$, canonical $(a,b,\dots)$ and ergodic $(\alpha, \beta,\dots)$ bases, respectively. Note that the computational basis is a basis for the full Hilbert space, whereas the canonical and ergodic are bases for the ground state sub-space, which we label $\mathcal{H}_{GS}$.

\subsection{Modular transformations: Dehn twists}\label{sec: Dehn twists}

Finally, we need to define the notion of a modular transformation for a microscopic system~\cite{CFTBook,Beverland2016,Zhang2012}. In analogy with a continuum TQFT, we define them to be the set of transformations that preserve the ground state manifold. More precisely, let $\mathcal{H}_\text{GS}$ be the Hilbert space spanned by the ground states, then \emph{an operator $\hat{M}$ corresponds to a modular transformation if $\hat{M}|\phi\rangle \in \mathcal{H}_\text{GS}$ for all $|\phi\rangle \in \mathcal{H}_\text{GS}$}.  The restriction of these operators to the ground state manifold are representations of the mapping class group (MPG) for the surface. For a torus, the MPG is the modular group generated by two elements, $s$ and $t$~\cite{CFTBook}. The first exchanges the vertical (meridian) and horizontal (longitude) non-contractible loops of the torus, see Fig.~\ref{fig: modular transformations}(c). The second is the Dehn twist, which corresponds to cutting the torus along a non-contractible curve, twisting one of the ends by a full rotation then gluing the torus back together, see Fig.~\ref{fig: modular transformations}(d). 

A given TQFT is defined in terms of how modular transformations, given by the modular $S$ and $T$ matrices, act on the Hilbert space $\mathcal{H}_{GS}$. In particular, with respect to the vertical canonical basis, for a non-chiral theory we have the matrix elements~\cite{SteveTopo,Beverland2016,CFTBook}
\begin{equation}\label{eq: modular S and T}
\begin{aligned}
    \langle b | \hat{S} | a \rangle &= S_{ab},\\
    \langle b | \hat{T} |a \rangle &= \theta_a \delta_{ab},
\end{aligned}
\end{equation}
where $S_{ab}$ contains the mutual statistics of anyons $a$ and $b$, and $\theta_a$ is the topological spin or exchange statistics for $a$, see Appendix.~\ref{ap: TQFT} for more details. The modular matrices are defined in terms of their action on the Wilson operators. Namely, the operator $\hat{T}$ preserves the ground states sub-space of $\hat{H}$ and transforms the Wilson loop operators as $\langle a | \hat{T} \hat{W}^v_c \hat{T}^\dagger | b\rangle = \langle a |\hat{W}^v_c | b\rangle$ and $\langle a | \hat{T} \hat{W}^h_c \hat{T}^\dagger | b\rangle = \langle a |\hat{W}^d_c | b\rangle$, where $\hat{W}^d_c$ is the diagonal Wilson operator shown in Fig.~\ref{fig: modular transformations}(d). In the following we focus on the Dehn twist since there are two inequivalent types, $\hat{T}$ and $\hat{T}'$, around the vertical and horizontal loops of the torus, as shown in Fig.~\ref{fig: modular transformations}(d). These two transformations also generate the modular group. We will consider the vertical (meridian) Dehn twist $\hat{T}$ and work in the vertical canonical basis throughout this paper, but the arguments can equally be repeated with for the horizontal (longitude).

\begin{figure}[t!]
    \centering
    \includegraphics[width = \columnwidth]{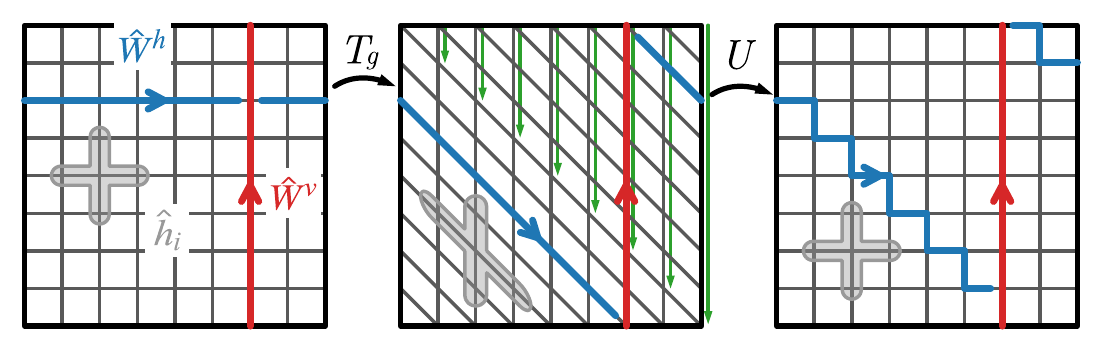}
    \caption{The microscopic implementation of the modular Dehn twist consisting of two parts. The operator $\hat{T}_g$ is a permutation of the computational basis states implementing a shear of the lattice. This causes a deformation of the lattice defined by $\hat{h}_i$ (shown in grey) and maps the horizontal Wilson loop to a diagonal one. The operator $\hat{U}$ is a local stoquastic adiabatic transformation that returns the lattice back to its original form while leaving the Wilson loops (red and blue lines) invariant. Note we show a square lattice of supersites with nearest neighbour interactions for simplicity but the procedure is independent of the microscopic details.}
    \label{fig: DehnTwist lattice}
\end{figure}

We next provide a model-agnostic microscopic protocol for implementing the $T$-matrix. This consists of two parts: a global geometric Dehn twist~\cite{Moradi2014,Mei2014,Zhu2018} implemented by the operator $\hat{T}_g$, and a local adiabatic change to the microscopic Hamiltonian $\hat{U}$, shown schematically in Fig.~\ref{fig: DehnTwist lattice}. The operator $\hat{U}$ fixes the local distortions introduced by $\hat{T}_g$, while not affecting any topological properties. 

\begin{figure*}[t!]
    \centering
    \subfigimg[height=.31\textwidth]{\textbf{(a)}}{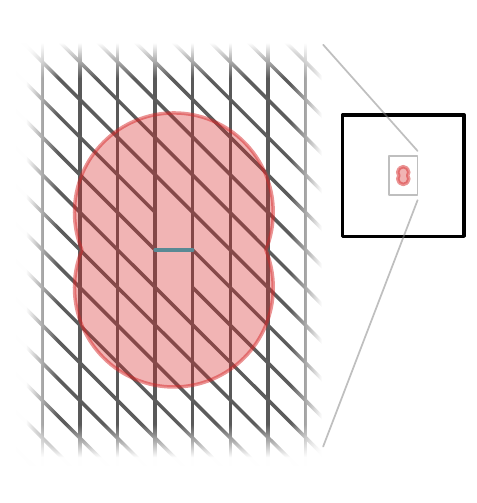}
    \subfigimg[height=.31\textwidth]{\textbf{(b)}}{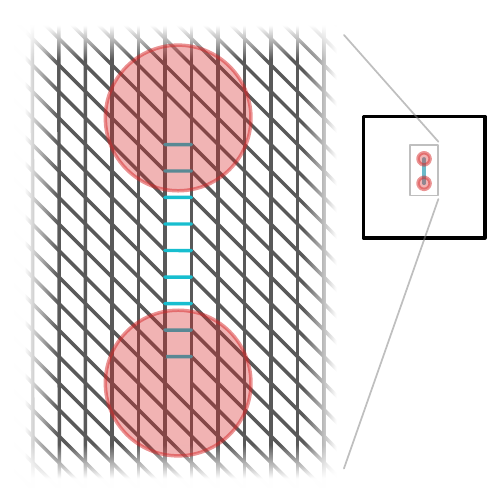}
    \subfigimg[height=.31\textwidth]{\textbf{(c)}}{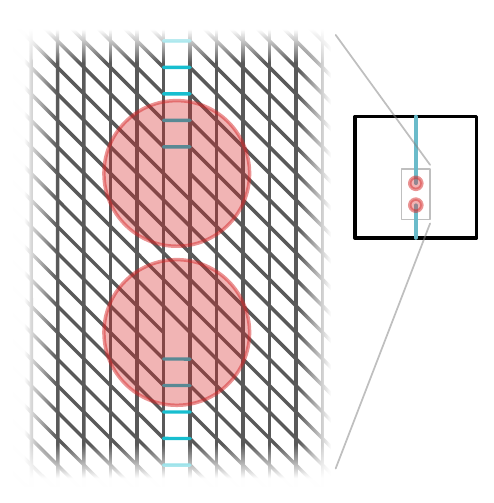}
    \caption{A step-by-step microscopic prescription for the LSAP that implements $\hat{U}$. \textbf{(a)} Two local lattice dislocations are created. This happens in a region of radius $R_h$ (shown in red) wherein the computational degrees of freedom are polarized. \textbf{(b)} The dislocations are slowly dragged apart. The red regions are polarized but everywhere else the Hamiltonian is the sum of terms $h_i$ that locally match the lattice connectivity. \textbf{(c)} The dislocations are dragged around the vertical non-contractible direction of the torus until they meet again, leaving behind a string of reconnected bonds (blue bonds).}
    \label{fig: LSAP}
\end{figure*}

Let us for the moment denote the microscopic computational basis (the basis $\{|i\rangle\}$ on which $\hat{H}$ stoquastic) using the notation $|\{s_{x,y}\}\rangle  = \bigotimes_{x,y}|x,y,s\rangle$, where $(x,y)$ denotes the (super-)site location and $s$ labels the internal degrees of freedom. The geometric Dehn twist corresponds to the map $\hat{T}_g |\{s_{x,y}\}\rangle = |\{s_{x,y-x}\}\rangle$, where addition is modulo $L$, which implements a right-handed Dehn twist around the meridian of the torus, see Fig.~\ref{fig: modular transformations}(b). The geometric Dehn twist is a permutation in the computational basis by definition and so it yields a new stoquastic Hamiltonian $\tilde{H} = \hat{T}_g \hat{H} \hat{T}_g^{\dagger}$ with a new connectivity structure. Since $\hat{H}$ yields a TQFT at long wavelength and since TQFTs partition functions are metric independent~\cite{Kaul2005}---in particular under the metric change induced by $\hat{T}_g$---$\hat{H}$ and $\tilde{H}$ are in the same phase of matter. Since the two Hamiltonians $\hat{H}$ and $\tilde{H}$ are in the same phase (and also have identical spectrum since $T_g$ is unitary), there exists an adiabatic path connecting the ground states of these Hamiltonians taking time $\mathcal{O}(L^0)$~\cite{Chen2010,Zeng2015a}. In fact, we show in Sec.~\ref{sec: LSAP} that there exists an adiabatic path that preserves the topological labelling of the ground states at all points and introduces no relative phases between them. We denote by $\hat{U}$ the unitary transformation obtained from this adiabatic path which we refer to as the local stoquastic adiabatic path (LSAP). Importantly, our prescription corresponds precisely to the definition of the modular $T$-matrix since it preserves the ground state manifold and transforms the Wilson loop operators in the correct way. Our goal is to show that given a stoquastic Hamiltonian, and the corresponding ergodic basis states $|\alpha\rangle$, the matrix elements $\langle\beta | \hat{T} |\alpha\rangle \geq 0$, are non-negative.


In the context of non-chiral and abelian theories our prescription for $\hat{T}$ provides a microscopic procedure to implement the modular transformation exactly. A similar procedure was considered in Ref.~\cite{Zhu2018}, however, their prescription for $\hat{U}$ was specific to string-net models. Furthermore, our procedure allows us to more carefully understand the action of the Dehn twists and keep track of the non-negative ground state basis. We will revisit the direct geometric approach when we consider chiral and non-abelian theories in Sec.~\ref{sec: non-commuting}.

\section{The local stoquastic adiabatic path}\label{sec: LSAP}

Next we precisely define the transformation $\hat{U}$, which relates the ground states of $\tilde{H}$ and $\hat{H}$, and prove that it acts as the identity map between the ground state manifold---that is, the labelling of the canonical and ergodic states is the same before and after the adiabatic evolution. Let us define this idea of an identity map more precisely. Since $\hat{H}$ and $\tilde{H}$ are in the same phase we can separately define a canonical basis $|a\rangle$ and $|\tilde{a}\rangle$ for each each in terms of vertical and horizontal Wilson loop operators. Furthermore, by stoquasticity, both Hamiltonians have an ergodic basis, labelled $|\alpha\rangle = \sum_b V_{\alpha b} |b\rangle$ and $|\tilde{\alpha}\rangle = \sum_{\tilde{b}} \tilde{V}_{\tilde{\alpha} \tilde{b}} |\tilde{b}\rangle$. \emph{The adiabatic evolution $\hat{U}$ is an identity map if $|a\rangle = \hat{U} | \tilde{a}\rangle$ and the matrices relating the canonical and ergodic bases are equal, i.e., $V = \tilde{V}$}. Note that such a map is not unique since it is only defined in terms of its action between the ground state sub-spaces. 

We will now provide an explicit adiabatic procedure for $\hat{U}$ interpolating between $\tilde{H}$ and $\hat{H}$ in Sec.~\ref{sec: the path}. We will show in Sec.~\ref{sec: adiabatic} that this adiabatic evolution indeed preserves the ground state sub-space and preserves the labelling of the canonical basis states. In Sec.~\ref{sec: imaginary} we will show that we can equivalently consider imaginary time evolution as a way of implementing the adiabatic path. Using the imaginary time evolution is important for allowing us to keep track of the non-negativity of the ergodic ground states. We need both the adiabatic and imaginary time evolution since the adiabatic evolution alone does not guarantee non-negativity of the ergodic states, whereas the imaginary time evolution alone does not guarantee that we do not leave the phase. In summary, the adiabatic evolution allows us to show that $\hat{U}$ is an identity map with respect to the canonical basis, and the imaginary time evolution ensures that $\tilde{V} = V$. Combining these two approaches we show that $\hat{U}$ is indeed an identity map as defined above and preserves the non-negativity of the ergodic states.

\subsection{Assumptions}\label{sec: assumptions}

In our proof we use a couple of reasonable assumptions, which we detail for the benefit of a mathematical audience. 

\emph{Assumption I}. There exist some finite burger's vector such that two lattice dislocation with such opposite burger's vectors do not change the ground-state degeneracy. Alternatively there exists stoquastic perturbation which does not drive a phase transition such that the previous statement holds. 

\emph{Assumption II}. The adiabatic time evolution along a time interval $\tau$ from $\hat{H}_{t-1}$ to $\hat{H}_{t}$ where the difference between $\hat{H}_{t-1}$ and $\hat{H}_t$ is strictly local, remains in the ground state multiplet to any desirable fidelity for large enough $\tau$. In addition the imaginary time evolution along a time interval $\tau$ from $\hat{H}_{t-1}$ to $\hat{H}_{t}$ projects to the ground state multiplet of $\hat{H}_{t}$ with any desirable fidelity for large enough $\tau$. Alternatively there exists a (potentially time dependent) stoquastic perturbation acting in the vicinity of the region where $\hat{H}$ and $\hat{H}'$ differ, such that the previous statements hold. 

\emph{Comments on assumptions}. Concerning Assumption I we stress that a TQFT does not imply a ground state degeneracy associated with a dislocation. Indeed dislocation is a form of torsion under which the theory is invariant. In Appendix~\ref{ap: flux binding} we show that such a degeneracy can be either be lifted by a stoquastic perturbation and incorporated into our procedure, or it would mean there is spontaneous breaking of stoquasticity. The latter outcome was proven impossible for the case of double degeneracy in Ref.~\cite{Ringel2017} and we believe to be impossible in general. If Assumption II fails, it means that the microscopic model exhibits arbitrarily slow relaxation for some local perturbations and arguably realizes a different phase of matter than that implied by the TQFT.

\subsection{The path}\label{sec: the path}

We begin by constructing a sequence of Hamiltonians $\hat{H}_t$, with $t=0,\ldots,L^2$, from $\hat{H} = \sum_i \hat{h}_i$ by inserting dislocations or moving them, as shown schematically in Fig.~\ref{fig: LSAP}. This is done as follows: all the lattice sites $j$ that are at least $R_H$-away from the dislocations result in an $\hat{h}_j$ term in $\hat{H}_t$, possibly with different connectivity to $\hat{H}$ or $\tilde{H}$; and sites $k$, associated with position $k_x,k_y$, which are nearer to the dislocations, do not result in an $\hat{h}_j$ operator but in a polarization operator $-\lambda |k_x,k_y,s\rangle \langle k_x,k_y,s|$ with $\lambda$ being some finite fixed number, see Fig.~\ref{fig: LSAP}. Notably, the resulting $\hat{H}_t$ is always stoquastic on the computational basis and differs from $\hat{H}_{t-1}$ by a local perturbation. We denote the Hilbert space of the ground state manifold at step $t$ as $\mathcal{H}_{GS,t}$.

Explicitly, $\hat{H}_0 = \tilde{H}$ and $\hat{H}_1$ is constructed from $\hat{H}_0$ but with two nearby dislocations, more accurately a dislocation and an anti-dislocation with a Burger's vector consistent with Assumption I. These dislocations can be initially on neighbouring sites, see Fig.~\ref{fig: LSAP}(a). Following this, $\hat{H}_2$ to $\hat{H}_{L-1}$ are associated with discretely moving one of the dislocations around a vertical loop of the torus by one vertical site at each $t$ increment. Once the dislocations are separated by a distance greater than $2R_H$ the terms in the Hamiltonian on sites between the dislocations are of the same form $\hat{h}_i$, however, they now match the new reconnected lattice, see Fig.~\ref{fig: LSAP}(b). We proceed to move the dislocations until they are again next to each other, where at $\hat{H}_{L}$ this pair is removed as it meets around the handle of the torus. The result is a change in the lattice structure where a column of diagonal bonds become horizontal. We next continue this process so that $\hat{H}_{L+1}$ introduces a dislocation pair at the nearby column, $\hat{H}_{L+2}$ till $\hat{H}_{2L-1}$ moves that second pair vertically around the torus and $\hat{H}_{2L}$ removes them. This continues with further pairs along nearby columns until $\hat{H}_{L^2}=\hat{H}$. It can be checked that this process maps the lattice underlying $\tilde{H}$ to that underlying $\hat{H}$.

\subsection{Adiabatic evolution}\label{sec: adiabatic}

\begin{figure}[t!]
    \centering
    \subfigimg[width=.43\columnwidth]{\textbf{(a)}}{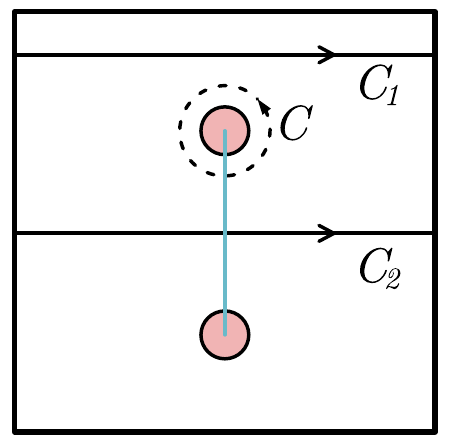}\qquad
    \subfigimg[width=.43\columnwidth]{\textbf{(b)}}{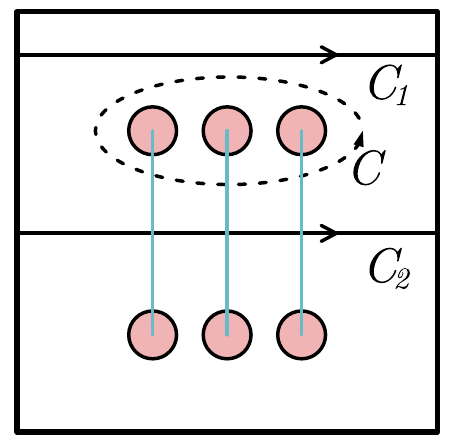}
    \caption{Schematic of the system with separated dislocations (red regions) with reconnected bonds in between (blue lines). \textbf{(a)} The case where dislocations bind no flux and so the state is an eigenstate of $\hat{\Omega}_1(C)$, meaning that the Wilson loops along $C_1$ and $C_2$ have the same action. \textbf{(b)} The case where dislocations bind a topological flux $a$. Since we treat abelian TQFTs, we can create $n$ such dislocation pairs with $a^n = 1$. In this case again the Wilson loop operators on  $C_1$ and $C_2$ agree.}
    \label{fig: flux binding}
\end{figure}

We split the adiabatic time evolution into steps, i.e., $\hat{U} = \lim_{\tau_1\rightarrow\infty}\prod_t \hat{U}_t(\tau_1)$, where each adiabatic evolution is implemented over a time $\tau_1$ by the operator $\hat{U}_t(\tau_1) = \mathcal{T}e^{-i\int_{0}^{\tau_1}d\tau (\tau \hat{H}_t + (\tau_1 - \tau) \hat{H}_{t-1})/\tau_1 + \delta(\tau/\tau_1)}$, which is a time-ordered exponential and $\delta(\tau/\tau_1)$ is some time-dependent stoquastic perturbation with support on the local region where $\hat{H}_{t-1}$ and $\hat{H}_t$ differ. This local stoquastic perturbation is included so as to ensure that the adiabatic evolution preserves to the ground state sub-space to any desirable fidelity for large enough $\tau$, as in Assumption II.

Let us consider the adiabatic evolution between steps $t-1$ and $t$. We proceed inductively and assume that $\hat{H}_{t-1}$ has a canonical basis states $|a,t-1\rangle$, which are labelled by the Wilson loop operators $\hat{W}^v_a$. Note that at a step $t$ where there are no dislocations we are free to move the vertical Wilson loop operator to act along any path that winds around the vertical direction of the torus. In particular we can choose this to be along a path at the far side of the torus, maximally far from the pair of dislocations that we create at the next step. While we move this particular pair of dislocations we can can keep this vertical path fixed and change it as necessary once we have annihilated this pair. We then define the states
\begin{equation}
    |a,t\rangle_{U} = \hat{U}_t(\tau_1) |a,t-1\rangle.
\end{equation}
This state $|a,t\rangle_{U}$ depends implictly on $\tau_1$ and differs from a ground state of $\hat{H}_t$ by an amount $\epsilon_1 = 1 - \max \{ |\langle \phi | a,t\rangle_{U}| \,:\, |\phi\rangle \in \mathcal{H}_{GS,t}\} $. Since $\hat{U}_t$ is an adiabatic evolution between $\hat{H}_{t-1}$ and $\hat{H}_{t}$, which differ only locally, the difference from a ground state quantified by $\epsilon_1$ can be made arbitrarily small by taking large enough $\tau_1$, via assumption II. 

We also have that, $\hat{W}^v_a \hat{U}_t(\tau_1) = \hat{U}_t(\tau_1)\hat{W}^v_a$. To see this, we can write $\hat{U}_t(\tau_1) = \mathcal{T} e^{-i \int^{\tau_1}_0 \hat{H}(\tau)}$, where $\hat{H}(\tau) = (\tau \hat{H}_t + (\tau_1 - \tau) \hat{H}_{t-1})/\tau_1 + \delta(\tau/\tau_1) = \hat{H}_{1} + \hat{H}_2(\tau)$ interpolates between $\hat{H}_{t-1}$ and $\hat{H}_t$. The time dependent piece $\hat{H}_2(t)$ only has non-trivial support on the finite region (radius $\sim R_H$) around the dislocation that we are moving by one lattice site. The time independent piece $\hat{H}_1$ is equal to $\hat{H}_{t-1}$ on all sites outside of that finite region. Since, the Wilson loop operator $\hat{W}_a^v$ does not have common support with $\hat{H}_2$ and commutes with $\hat{H}_1$---since we chose the Wilson loop to act on a path on the far side of the torus, far from the dislocations---we have that $\hat{W}^v_a \hat{H}(\tau) = \hat{H}(\tau) \hat{W}^v_a$. Furthermore, $\hat{W}^v_a$ commutes with each term in the time-ordered Taylor expansion of $\hat{U}_t(\tau_1)$, namely,
\begin{equation}\label{eq: adiabatic Taylor}
    \int^{\tau_1}_0 \int^{\tau_2}_0  \cdots \int^{\tau_{n-1}}_0 \hat{H}(\tau_2) \hat{H}(\tau_3) \cdots \hat{H}(\tau_n) \;\text{d}\tau_n \cdots \text{d}\tau_2,
\end{equation}
where $\tau_1 > \tau_2 > \cdots > \tau_{n-1}$. Since $\hat{W}^v_a$ commutes with each terms of the form in Eq.~\eqref{eq: adiabatic Taylor} for each order $n$ and for each value of $\tau_1$, we also have the stronger statement that that $[\hat{W}^v_a, \hat{U}_t(\tau_1)] = 0 $, independent of $\tau_1$. Since $\hat{\Omega}^v_a |b, t-1\rangle = \delta_{a,b} |b, t-1\rangle$, we therefore also have that $\hat{\Omega}^v_a |b, t\rangle_{U} = \delta_{a,b} |b, t\rangle_{U}$ and so these states correspond to the canonical basis states for $\hat{H}_t$.

In summary, there still remains the possibility that the evolution induces state dependent phases, i.e., $|a,t\rangle_U = e^{i\phi_a(\tau_1)} |a,t\rangle + \epsilon_1$. This freedom in the relative phase can be fixed by keeping track of the action of the horizontal Wilson loop operators. Unlike the vertical Wilson loop operators, we will have to make a different choice of the horizontal loop operators, which are used the generate the canonical states, at a certain point while dragging the dislocations. We therefore need to relate those horizontal Wilson loop operators outside of and between the dislocations. We then have to split into two possibilities, depending on whether or not the dislocations bind a topological flux.

Let us first examine the case that the dislocations \emph{do not} bind a flux, shown in Fig.~\ref{fig: flux binding}(a). Consider a step $t$ such that the dislocations are far enough apart that we can take the horizontal paths $C_1$ and $C_2$ and around and between the dislocations, such that we have well-defined Wilson loops that commute with the Hamiltonian, see Fig.~\ref{fig: flux binding}(a). Let $C$ be a curve surrounding one of the dislocations sufficiently far away from the polarized region. Not binding flux is then the statement that $\hat{\Omega}_1(C) |\phi\rangle = |\phi\rangle$, for all $|\phi\rangle \in \mathcal{H}_{GS,t}$. This allows us to freely bring the Wilson loops across this region without changing their expectation values (shown in appendix appendix~\ref{ap: TQFT}) and so $\hat{W}_a(C_1)|a,t\rangle = \hat{W}_a(C_2)|a,t\rangle$. We are able to generate the basis states using any horizontal Wilson loop operator that avoids the polarized regions, e.g., $|a,t\rangle_{U} = \hat{W}_a(C_1) |\mathds{1},t\rangle_{U} = \hat{W}_a(C_2) |\mathds{1},t\rangle_{U}$. Even for a system with polarized regions and with finite $\tau_1$, this defines an orthonormal set of states, see appendix~\ref{ap: orthogonality}. Since $\hat{W}_a(C_1)$ and $\hat{W}_a(C_2)$ both commute with $\hat{U}_t(\tau_1)$, this fixes the relative phase between the states.

In the second case, that the dislocations \emph{do} bind fluxes, let us restrict ourselves to consider the case of a unique anyon flux and the possibility of superpositions will be considered in Appendix~\ref{ap: flux binding}. Here we use the fact that the anyons of an abelian TQFT have an abelian group structure, and in particular each element $a\in \mathbb{A}$ has a finite order. That is, there exists finite $n$ such that $a^n = a \times a \times \cdots \times a = 1$. Therefore if a dislocation binds an anyon of order $n$ we can modify our procedure to create $n$ dislocations and move them around the lattice in parallel, as shown in Fig.~\ref{fig: flux binding}(b). These dislocations and paths should be separated by at least $2R_H$ such that they are independent and equivalent, but they can still be created and collectively moved by local (finite-ranged $\sim\mathcal{O}(n R_H)$) perturbations. In this case, the region containing the $n$ dislocations now has no total anyonic flux and we can again take a curve $C$ surrounding all $n$ such that $\hat{\Omega}_1(C) |b,t\rangle = |b,t\rangle$ for all $b$, see Fig.~\ref{fig: flux binding}(b), and again we fix the relative phase between the states.

By induction we therefore have that $\hat{U}$ preserves the ground state sub-space and preserves the labelling of the canonical basis states. However, we still need a handle on the matrix $V$ relating the canonical and ergodic basis states at each step $t$.

\subsection{Imaginary time evolution}\label{sec: imaginary}

Next let us consider imaginary time evolution along the same Hamiltonian path. We do this so that we can keep track of the non-negativity of the ergodic basis states and of the matrix $V$ relating these to the canonical basis states. The imaginary time evolution operator between each step is $\hat{P}_t(\tau_2) = N^{-1} e^{-\tau_2 \hat{H}_t}$, where $N = |e^{-\tau_2 \hat{H}_t}|a,t-1\rangle|$, and $|a,t-1\rangle$ is any ground state at time $t-1$---we will show that this operator is well defined below. We will now show that the restriction to the ground state sub-space of the unitary adiabatic evolution and the imaginary time evolution are equivalent. More precisely, we define the restriction of an operator $\hat{O}$ to the ground state manifold between steps as 
\begin{equation}
    [\hat{O}]_{t,t-1} = \sum_{|a,t\rangle, |b,t-1\rangle} |a,t\rangle \langle a,t| \hat{O} |b,t-1\rangle \langle b,t-1 |.
\end{equation}
We will show that 
\begin{equation}\label{eq: U and P equivalence}
    [\hat{U}_t(\tau_1)]_{t,t-1} = e^{i\phi(\tau_1)} [\hat{P}_t(\tau_2)]_{t,t-1} + \epsilon \hat{R},
\end{equation}
where $R_{ab} = \langle a,t| \hat{R} |b,t-1\rangle$ is a matrix with elements $|R_{ab}|\leq 1$ and $\epsilon$ can be made arbitrarily small by taking larger values of $\tau_1, \tau_2$. We will show that $\hat{P}_t(\tau_2)$ is non-negative and so the phase factor $e^{i\phi(\tau_1)}$ is dependent only on $\tau_1$ and can be incorporated into $\hat{U}_t(\tau_1)$.

Again we proceed inductively and consider the states defined under imaginary time evolution between steps $t-1$ and $t$, that is,
\begin{equation}
    |a,t\rangle_{P} = \hat{P}_t(\tau_2) |a,t-1\rangle,
\end{equation}
which implicitly depends on $\tau_2$. The normalization in $\hat{P}_t(\tau_2)$ is well defined because by assumption the state at time $t-1$ can be generated by a horizonatal Wilson loop operator, i.e., $|a,t-1\rangle = \hat{W}^h_a|1,t-1\rangle$. We can, however, choose this operator to be far from the lattice dislocations such that $[\hat{W}^h_a, \hat{H}_t] = 0$, and so we have
\begin{equation}
\begin{aligned}
    \langle a,{t-1} | e^{-2\tau_2 H_t} | a,t-1 \rangle &= \langle 1, {t-1} | \hat{W}^h_{\bar{a}} e^{-2\tau_2 H_t} \hat{W}^h_{a}| 1, t-1 \rangle \\
    &= \langle 1,{t-1} | e^{-2\tau_2 H_t} \hat{W}^h_{\bar{a}} \hat{W}^h_{a}| 1,t-1 \rangle \\
    &= \langle 1,t-1 | e^{-2\tau_2 H_t}| 1,t-1 \rangle \equiv N^2,
    \end{aligned}
\end{equation}
independent of $a$.
The states $|a,t\rangle_{P}$ differ from ground states of $\hat{H}_t$ by $\epsilon_2 = 1 - \max \{ |\langle \phi | a,t\rangle_{P}| \,:\, |\phi\rangle \in \mathcal{H}_{GS,t}\} $. Using Assumption II, we can take $\epsilon_2$ arbitrarily small by increasing $\tau_2$. Furthermore, since $[\hat{W}^v_a, \hat{H}_t] = 0$ we have that $\hat{\Omega}^v_a |b,t\rangle_P = \delta_{a,b} |b,t\rangle_P$ and so these are the also canonical basis states for $\hat{H}_t$, with the same labelling as those generated by $\hat{U}_t$. We can then use similar arguments as for the adiabatic evolution for the horizontal Wilson operators---possibly needing to modify our procedure as before to account for dislocations binding flux---such that $|a,t\rangle_P = \hat{W}_a(C_1)|1,t\rangle_P = \hat{W}_a(C_2) |1,t\rangle_P$. We therefore have that $|a,t\rangle_{U} = e^{i\phi(\tau_1,\tau_2)} |a,t\rangle_{P} + \mathcal{O}(\max(\epsilon_1,\epsilon_2))$, related by a global but not state dependent phases, and where $\epsilon_1$ and $\epsilon_2$ can be made arbitrarily small by taking large enough $\tau_1,\tau_2$.

Importantly we also have that the imaginary time evolution maps ergodic basis states to ergodic basis states. Indeed, $e^{-\tau_0 \hat{H}_t}=e^{-\tau \hat{H}_t+\Lambda \hat{\mathds{1}}}e^{-\Lambda \hat{\mathds{1}}}$ and by virtue of $\hat{H}_t$'s stoquasticity there exist a large enough $\lambda$ such that $[-\tau \hat{H}_t+\Lambda \hat{\mathds{1}}]_{ij}$ is element-wise non-negative. Using a Taylor expansion of $e^{-\tau \hat{H}_t+\Lambda \hat{\mathds{1}}}$ and the fact that non-negative matrices are closed under addition and multiplication one finds that $e^{-\tau \hat{H}_t+\Lambda \hat{\mathds{1}}}$ is element-wise non-negative and similarly that $e^{-\tau \hat{H}_t+\Lambda \hat{\mathds{1}}}e^{-\Lambda \hat{\mathds{1}}}$ is element-wise non-negative. Consequently we find that our imaginary time evolution maps non-negative state to non-negative states. Therefore the relative phase between $|a,t\rangle_U$ and $|a,t\rangle_P$ is set only by $\tau_1$ and can be incorporated into the adiabatic evolution by considering the phase of the overlap of these two states. This establishes that the adiabatic and imaginary time evolution, as defined, have the same action on the ground state sub-spaces along our path.

\subsection{$\hat{U}$ is an identity map}\label{sec: non-negative}

We now have all of the machinery necessary to show that the operator $\hat{U}$ is an identity map. Using the imaginary time evolution we have shown that each $\hat{U}_t$ maps non-negative states to non-negative states. Therefore $\hat{U}$ is generally a permutation between the ergodic basis states of $\tilde{H}$ and $\hat{H}$. However, since $\hat{U}_t$ also preserves the canonical basis states at each step, we also have that 
\begin{equation}
\begin{aligned}
    |\alpha,t\rangle &= \lim_{\tau_2 \rightarrow \infty } \hat{P}_t(\tau_2) |\alpha, t-1\rangle = \lim_{\tau_1 \rightarrow \infty } \hat{U}_t(\tau_1) |\alpha, t-1\rangle \\
    &= \lim_{\tau_1 \rightarrow \infty } \hat{U}_t(\tau_1) \sum_b V_{\alpha b} |b,t-1\rangle = \sum_b V_{\alpha b} |b,t\rangle,
\end{aligned}
\end{equation}
that is, the relationship between ergodic and canonical states, given by the unitary matrix $V_{\alpha b}$, is the same at time $t$ as it was at time $t-1$. Therefore we have that $\hat{U}_t = \lim_{\tau_1\rightarrow\infty}\hat{U}_t(\tau_1)$ is an identity map between the ergodic basis states, i.e., $|\alpha,t\rangle = \hat{U}_t |\alpha,t-1\rangle$. Since this holds for each $t$ we have our result that $\hat{U} = \prod_t \hat{U}_t$ is an identity map, and $|\alpha\rangle = \hat{U} |\tilde{\alpha}\rangle$.

\section{$T$ and $S$ are non-negative}\label{sec: T non-negative}

In the previous section we showed that $\hat{U}$ is an identity map and preserves the expectation values of Wilson loop operators around the two handles of the torus. Combined with the action of the geometric Dehn twist, we see that $\hat{T}$ preserves the ground state sub-space of $\hat{H}$ and transforms the Wilson loops in the required manner, i.e.,
\begin{equation}
\begin{aligned}
\langle a | \hat{T} \hat{W}^h_c \hat{T}^\dagger | b\rangle = \langle \tilde{a} | \hat{T}_g \hat{W}^h_c \hat{T}_g^\dagger | \tilde{b} \rangle =  \langle a |\hat{W}^d_c | b\rangle,\\
\langle a | \hat{T} \hat{W}^v_c \hat{T}^\dagger | b\rangle = \langle \tilde{a} | \hat{T}_g \hat{W}^v_c \hat{T}_g^\dagger | \tilde{b} \rangle =  \langle a |\hat{W}^v_c | b\rangle,
\end{aligned}
\end{equation}
confirming that $\hat{T}$ implements the modular Dehn twist.
We are now in a position to show that $\hat{T}$ is non-negative in the ergodic basis. This shows that there must exist a basis under which the $T$-matrix of the underlying TQFT can be made non-negative, and further that there must exist a basis for which both and $S$ and $T$ are non-negative.

Firstly, $\langle \tilde{\alpha} | \hat{T}_g | \beta\rangle \geq 0$ since $\hat{T}_g$ is non-negative in the computational basis and therefore maps non-negative states to non-negative states. Combined with the fact that $\hat{U}$ is an identity map between the ergodic bases, this means that $\hat{T} = \hat{U}\hat{T}_g$ is in general a permutation of the ergodic basis states, i.e.,
\begin{equation}
    \langle \alpha |\hat{T} |\beta\rangle = \sum_{a} V^{}_{\alpha a} \theta^{}_a V^\dag_{a\beta} =  P_{\alpha \beta} \geq 0.
\end{equation}
The eigenvalues of a permutation matrix necessarily form complete sets of roots of unity (as shown in Appendix.~\ref{ap: permutations}), which establishes our result.

Since our choice of vertical Dehn twist was arbitrary we could have similarly considered horizontal Dehn twists with the same result, that is, there exists a basis---the ergodic basis---in which $\hat{T}$ and $\hat{T}'$ are simultaneously non-negative permutation matrices. By extension, since non-negative matrices are closed under multiplication, in this basis the $S$ matrix is also non-negative, i.e.,
\begin{equation}
     \langle \alpha |\hat{S} | \beta\rangle = \sum_{a,b} V_{\alpha a} S_{ab} V^\dag_{b\beta} = P'_{\alpha \beta} \geq 0,
\end{equation}
for some other permutation matrix $P'_{\alpha\beta}$.

In summary, if the Hamiltonian is locally stoquastic and therefore has a non-negative ergodic ground state basis, then with respect to this ergodic basis the $S$ and $T$ matrices of the underlying TQFT must be non-negative permutation matrices. Indeed, in this basis any modular matrix must be non-negative since non-negative matrices are closed under multiplication. Contrapositively, if there does not exist any unitary transformation for which the $S$ and $T$ matrix are non-negative, then the Hamiltonian cannot be made stoquastic by any local transformation and there is an \emph{intrinsic sign problem}. A simpler sufficient test for intrinsic sign problems follows from the fact that such a non-negative $T$ matrix must be a permutation matrix by unitarity and so its eigenvalues necessarily come in complete sets of roots of unitary. Thereby, by simply looking at the topological spins in the theory we can potentially diagnose an intrinsic sign problem. Similar to the fermion sign-problem, this has the appealing interpretation that the intrinsic sign problem is intimately linked to the statistics of the excitations in the theory.

\section{Generalisations}\label{sec: general}

Above we went through the trouble of carefully defining $\hat{U}$ and showing that is corresponded to a identity map. This used the exact commutation of the Wilson loop operators and the Hamiltonian and the fact we have an exact degeneracy. This, however, was not crucial to our arguments and we can also relax these conditions to cover a more general set of Hamiltonians. 

The procedure for $\hat{U}$ also relied on us restricting to abelian non-chiral models. Chirality and non-abelian statistics introduce subtleties into the procedure for $\hat{U}$---regarding the dislocations binding topological flux---that go beyond the scope of this paper. However, using conjectured results from the literature that have been analytically and numerically verified, we can extend our results to cover chiral and non-abelian models as well.

\subsection{Beyond strictly commuting Wilson operators}\label{sec: non-commuting} 

So far we have focused on the case of strictly local and perfectly commuting Wilson operators which imply various neat properties such as an exact ground state degeneracy. This behavior is, however, fine-tuned, since any generic perturbation would split the exact ground state degeneracy by some exponentially small factor in the system size ($\eta \approx e^{-L}$). This in turn also implies that the Wilson operator cannot commute with the Hamiltonian and obey the Verlinde algebra exactly as those two properties would imply an exact degeneracy. 
Here we show how to extend our proof into a physical argument that is valid in this more generic setting where (a) The Wilson operator, which previously had finite support around their path, are now allowed to have exponentially decaying tails. (b) The commutation relation between the Wilson operator and the Hamiltonian may be non-zero but exponentially small in system size. We shall now argue that by letting $\tau_1$ scale exponentially with the system size ($L$), $\tau_2$ to scale polynomially with $L$, and using our freedom in choosing $L$ as large as needed, we can make the above reasoning accurate even in this more complicated setting. 

Let us revisit the main arguments of the previous section with these changes in mind. Considering the adiabatic evolution between $\hat{H}_{t-1}$ and $\hat{H}_{t}$, it would now slightly mix the canonical states for two reasons: First the above two Hamiltonians and therefore $\hat{U}_t(\tau_1)$, do not perfectly commute with the Wilson operators. Second even if we had not changed the Hamiltonian at all, the ground-state superpositions would change when $\tau_1$ becomes comparable to the inverse energetic splitting of the ground state multiplet. To control this latter issue its sufficient to take $\tau_1 \ll \eta^{-1}$, say $\tau_1 = \eta^{-1/2}$, thereby making this discrepancy exponentially small at large $L$. Considering the first issue, let us split $\hat{U}_t(\tau_1)$ to a product of $M$ short time evolutions $\Pi^{M}_{i=1} \hat{U}_i$. For large enough $M$, one has that $[\hat{W},\hat{U}_i]=\frac{\tau_1}{M}O_i \eta$, where $O_i$ some operator with norm (highest eigenvalue) of order $1$, localized to where the defects are moving and $\hat{W}$ is some Wilson operator. Hence commuting $\hat{W}$ through all $\hat{U}_i$ and using $... \hat{W} \hat{U}_i... = ...[\frac{\tau_1}{M}O_i \eta \hat{W}^{\dagger}+\hat{U}_i] \hat{W}...$ would give corrections of the order $\sum^M_{i=1} \frac{\tau_1}{M}|O_i||\hat{W}^{\dagger}| \eta+\sum_{i,j=1}^M \frac{\tau_1^2}{M^2}|O_i| |O_j||\hat{W}^{\dagger}||\hat{W}^{\dagger}| \eta^2+...$. Notably we used the fact that $\hat{U}_i$ are all unitaries and therefore do not affect operator norms. Given this and our choice for $\tau_1$, this second discrepancy is dominated by the first term scaling as $\tau_1 \eta=\eta^{1/2} \propto e^{-L/2}$. 

Next we consider the imaginary time evolution $\hat{P}(\tau_2)$. Again two similar issues arise: $\hat{P}(\tau_2)$ does not exactly preserve the ground states and it does not exactly commute with $\hat{W}$. The first issue generates a discrepancy similar to the previous case, however the second is more severe here: Indeed splitting $\hat{P}(\tau_2)$ into a product of $M$ short imaginary time evolutions, one would again have that $\hat{W}$ almost commutes with each of them with similar discrepancies. However whereas previously the small discrepancies got multiplied from the left and right by unitary matrices, now they'll be multiplied by imaginary time evolutions which may strongly affect the norm of these operators. For instance, say that $\hat{W}$ has a matrix element, proportional to $\eta$, between a ground state and an excited state at energy $\Delta$. Its imaginary-time conjugated version, $\hat{P}(\tau_2) \hat{W} \hat{P}(\tau_2)$, would have such a matrix element proportional to $\eta e^{\tau_2 \Delta}\propto e^{-L + \tau_2 \Delta}$. To keep such matrix elements under control, it is sufficient to take $\tau_2$ to scale as $\tau^{1.1}_2 = \frac{L}{2\Delta}$. As a consequence $\hat{P}(\tau_2)$ would now project on the ground-state subspace, with some exponentially small corrections which can be made to effectively vanish given our freedom in choosing $L$.

\subsection{Chiral and non-abelian models}\label{sec: chrial}

In order to go beyond the non-chiral and abelian models considered above, we rely on the following conjecture in Ref.~\cite{Moradi2014} for the universal wave-function overlap. If we define $\hat{T}_g$ and $\hat{S}_g$ as the operators that implement the maps $t: (x,y) \rightarrow (x,y-x)$ and $s: (x,y)\rightarrow(y,-x)$ in the computational basis, respectively, then
\begin{equation}\label{eq: Wen modular}
\begin{aligned}
    \langle a | \hat{T}_g | b\rangle = e^{-\alpha_TL^2 + \mathcal{O}(1/L^2)} T_{ab},\\
    \langle a | \hat{S}_g | b \rangle = e^{-\alpha_SL^2 + \mathcal{O}(1/L^2)} S_{ab},
\end{aligned}
\end{equation}
where $|a\rangle$ are the canonical basis states (i.e. MES~\cite{Zhang2012}), and $T_{ab} = e^{-2\pi i c_-/24}\theta_a \delta_{ab}$ and $S_{ab} = \mathcal{D}^{-1}\sum_c N^c_{ab} \frac{\theta_c}{\theta_a \theta_b} d_c$. Here $c_-$ is the chiral central charge, which is an additional topological invariant assumed to vanish in previous sections. Comparing with our procedure for implementing the modular transformations, omitting our $\hat{U}$ leads an exponential suppression factor determined by non-universal parameters $\alpha_T$ and $\alpha_S$. This suppression is due to the microscopic lattice distortions induced by the direct geometric transformations.  If $T_g (S_g)$ happens to be a symmetry of $\hat{H}$, then $\alpha_T=0$ $(\alpha_S = 0)$. The conjecture in Eq.~\eqref{eq: Wen modular} has been verified numerically and analytically in a large number of examples in Refs.~\cite{Mei2014,Moradi2014,Moradi2015a,He2014,Mei2017}.

If we assume that our Hamiltonian is stoquastic (possibly after a local unitary transformation) and has non-negative ergodic basis states, then since $\hat{T}_g$ and $\hat{S}_g$ are non-negative permutations in the computational basis, we can immediately see that we have
\begin{equation}
    \langle\alpha| \hat{T}_g |\beta\rangle \geq 0, \qquad \langle \alpha | \hat{S}_g | \beta \rangle \geq 0.
\end{equation}
We argue that this implies that $T_{\alpha\beta} \geq 0$ and $S_{\alpha\beta} \geq 0$. Let us focus on $T_{\alpha\beta}$. If $T_{\alpha\beta}=0$ then there is nothing to show. For $T_{\alpha\beta}\neq0$, we can normalise Eq.~\eqref{eq: Wen modular} to get
\begin{equation}\label{eq: fix alpha}
    1 = \frac{\langle \alpha | \hat{T}_g | \beta\rangle}{|\langle \alpha | \hat{T}_g | \beta\rangle|} = e^{-i \text{Im} (\alpha_T) L^2 + \mathcal{O}(L^{-2})} \frac{T_{\alpha\beta}}{|T_{\alpha\beta}|}.
\end{equation}
Since the factor $T_{\alpha\beta}/|T_{\alpha\beta}|$ is independent of $L$, Eq.~\eqref{eq: fix alpha} implies that $\text{Im}(\alpha_T) = 0\; (\text{mod } 2\pi)$ and therefore that $T_{\alpha\beta} >0$. The same arguments show that $S_{\alpha\beta} \geq 0$. 

Therefore we can generalise our results from the previous sections. Since $T_{\alpha\beta}$ and $S_{\alpha\beta}$ are unitary matrices and non-negative they are necessarily permutation matrices, and so we have the condition on their spectra. In particular, the modular matrix $T_{\alpha \beta}$ is a non-negative permutation matrix in the ergodic basis and so its eigenvalues $e^{-2\pi i c_-/24}\theta_a$ form complete sets of roots of unity. Since 1 must be contained within these sets, we have the simpler criterion that the Hamiltonian being stoquastic implies there exists $\theta_a$ such that $\theta_a = e^{2\pi i c_-/24}$. This is consistent with the results in Refs.~\cite{Ringel2017,Golan2020}. The equations~\eqref{eq: Wen modular} also apply for non-abelian theories, and so by using these conjectures we obtain the same results for non-abelian and chiral models. Namely, that the $T$ matrix has eigenvalues that form complete sets of roots unity and that the $S$ and $T$ matrix can simultaneously be made non-negative.

As an example consider a theory of Ising anyons which has topological spins $\theta_a \in \{1,-1,e^{2\pi i C_1/16} \}$, where $C_1$ is an odd integer~\cite{Kitaev2003}. This model is both non-abelian and chiral. Regardless of the value of the chiral central charge, we can easily see that $e^{-2\pi i c_-/24}\theta_a$ don't form complete sets of roots of unity. For all odd values of $C_1$ a Hamiltonian realising this anyon theory should have an intrinsic sign problem. Additionally, we have found numerically that the first one thousand $S\!U(2)_k$ TQFTs, as well as the bosonic Laughlin theories with $\nu = 1/Q$ for $Q\in 2\mathbb{N}$ up to 1000, all have intrinsic sign problems by these criteria.

\section{Discussion}\label{sec: discussion}

In this paper we have shown that a two-dimensional stoquastic (sign-problem-free bosonic) Hamiltonian is only compatible with certain TQFTs at low energy. Specifically, the topological spins must form complete sets of roots unity. We also showed a more restrictive condition that the $S$ and $T$ matrices of the TQFT must be such that they can be made simultaneously non-negative by a unitary transformation. This allows us to easily diagnose intrinsic sign problems by noting that if the parent Hamiltonian fails these criteria then it has an intrinsic sign problem that can't be removed by local unitary transformations. Similar to the fermion sign-problem, this has the appealing interpretation that the intrinsic sign problem is intimately linked to the statistics of the excitations in the theory.

While we have proven these statements for abelian non-chiral TQFTs, to tackle the more general problem of chiral non-abelian theories we had to rely on conjectured results. We expect that our arguments can be extended, but there are additional subtleties to be taken into account. Specifically, these concern the binding of topological flux to lattice dislocations. While for abelian theories we were able to simply modify our procedure in a stoquastic manner to deal with this possibility, this does not directly translate to non-abelian and chiral theories. We therefore need additional arguments to arrive at a non-negative procedure for implementing the modular transformations. Furthermore, in this paper we considered bosonic Hamiltonians. We do, however, conjecture that the results we have obtained are general and also apply to fermionic Hamiltonians---where being stoquastic is no longer the relevant condition for being sign problem free, see Ref.~\cite{Golan2020}.

In models with non-abelian excitations we additionally conjecture that all such topologically ordered phases have intrinsic sign problems. This conjecture is, however, supported by the fact that TQFTs capable of topological computation such as Ising anyons~\cite{Kitaev} and the double fibonnaci model~\cite{Levin2005} have intrinsic sign problems by our criteria. Indeed, all non-abelian theories that we have checked fail our criteria including those in Kitaev's 16-fold way~\cite{Kitaev}, the list of modular tensor categories with rank $\leq 4$ found in Ref.~\cite{Rowell2009}, and $S\!U(2)_k$ models (which we have checked numerically for $k$ up to $1000$)~\cite{Bonderson2007}. It remains an open question whether non-abelian fusion rules necessarily imply that the $S$ and $T$ matrices fail the criteria we have presented. 

That non-abelian models have intrinsic sign problems is perhaps expected and reassuring since several are capable of \emph{universal} quantum computation, and it is generally assumed that the complexity class BQP contains problems outside of P. However, we also note that some of the theories that are not computationally universal are nonetheless numerically hard. For instance, the double semion model, the Ising anyon model, and $S\!U(2)_k$ for $k=1,2,4$ all have intrinsic sign problems but are not computationally universal~\cite{Nayak2008}. Such models may therefore deserve further study from a complexity theory perspective for use as computational resources.

While our results imply that intrinsic sign problems are widespread amongst topologically ordered phases, there are important physical systems that do not suffer from them. Most notably, high-temperature superconductivity is believed to be due to non-chiral d-wave pairing, which is not precluded by our results and may be accessible using sign-free determinental Quantum Monte Carlo~\cite{Berg2019}. As another example, the spin liquid ground state of the frustrated Kagome Heisenberg antiferromagnet is currently believed to have $\mathbb{Z}_2$ topological order~\cite{Yan2011,He2017}, which admits a sign-free QMC representation. A wide class of symmetry protected topological phases also fall outside of the phases considered in this paper, and are believed to be sign-problem-free~\cite{Geraedts2013,Bondesan2017,Gazit2016}. 

The presence of an intrinsic sign problem also does not necessarily discount practical solutions to relevant problems in many-body physics. For instance, while some sign problems may be impossible to remove fully, they can come in varying severity. That is, despite the sign problem, it may still be possible to access large enough systems to extract the relevant physics. Moreover, there are several recent works focussing on \emph{easing} the sign problem to bring a larger set of problems within the reach of current technology~\cite{Klassen2019,Hangleiter2019}. The application of machine learning techniques has also found success beyond QMC for systems with a fermionic sign problem~\cite{Broecker2017,Liu2017}.

In this paper we have revealed fundamental obstructions to numerical simulations and constraints on phases realised by stoquastic Hamiltonians. We have introduced a new geometric viewpoint and new analytical tools to precisely study topological properties of microscopic Hamiltonians and procedures to manipulate them in a sign-free manner. We hope that this work can not only provide deeper insights into complexity classes and the difficulties of numerical simulations, but also inspire new approaches to study complex quantum many-body systems.

\begin{acknowledgements}
We are grateful for enlightening discussions with Dmitry Kovrizhin, Steve Simon, Paul Fendley, Clement Delcamp and Robert K{\"o}nig. We are particularly thankful to Dmitry Kovrizhin and Clement Delcamp for feedback on drafts of this paper. A.S. was funded by the European Research Council (ERC) under the European Union's Horizon 2020 research and innovation programme (grant agreement No. 771537). Z.R. was supported by the Israel Science Foundation (ISF grant 2250/19), the Deutsche Forschungsgemeinschaft (DFG, CRC/Transregio 183, EI 519/7-1), and the European Research Council (ERC, Project LEGOTOP).
\end{acknowledgements}

\hfill\\
\PRLsep

\appendix

\begin{center}
\textbf{APPENDICES}
\end{center}

\section{Results from linear algebra}

Here we present a few linear algebra results that are used in the main text. In particular we use a particular form of the Frobenius-Perron theorem as well properties of permutation matrices.

\subsection{Frobenius-Perron}\label{ap: Frobenius-Peron}

We wish to show that if we have a non-negative Hamiltonian $H$ with a degenerate ground state sub-space (here meaning the eigenspace with \emph{largest} eigenvalue), then there exists a basis for this sub-space that is strictly non-negative. We call this basis the \emph{ergodic} basis. This follows from the Frobenius-Perron theorem, which we will present for completeness.

First, a matrix $A$ is \emph{irreducible} if it cannot be conjugated into upper triangular form by a permutation $PAP^{-1}$. For example $\sigma^x$ is irreducible. Any \emph{reducible} square matrix $A$ can be reduced to upper triangular form by a permutation $P$, i.e., 
\begin{equation}
    P A P^{-1} = \left(\begin{array}{cccc}
    A_1 & * & * & \cdots \\
    0 & A_2 & * &  \\
    0 & 0 & A_3 &  \\
    \vdots & & & \ddots
    \end{array}\right )
\end{equation}
where each $A_i$ is irreducible.

\hfill\\\noindent\textbf{Frobenius-Perron Theorem for non-negative matrices:}\\
Let $A$ be a non-negative irreducible matrix with spectral radius $\rho(A) = r$, then:
\begin{itemize}[noitemsep,topsep=1pt]
    \item $r$ is a positive eigenvalue of $A$, and is simple and unique.
    \item the eigenvector corresponding to $\lambda = r$ is positive.
\end{itemize}
\hfill

\noindent\textbf{Corollary:} Let $H$ be a non-negative Hermitian matrix with $N$ degenerate ground states (largest eigenvalues), then
\begin{itemize}[noitemsep,topsep=1pt]
    \item there exists a non-negative basis $\{|\psi_1\rangle, \ldots, |\psi_N\rangle\}$ for the ground state manifold
    \item these non-negative ground states correspond to distinct ergodic sectors, i.e. $\langle \psi_i | k \rangle \langle k | \psi_j \rangle $, for each $i, j$ and for each $|k\rangle$ in the computational basis.
\end{itemize}

\noindent\textbf{Proof:} $H$ is reducible since if it were irreducible the ground state would be unique. Hence, there exists a permutation $P$ such that 
\begin{equation}
    P H P^{-1} = \left(\begin{array}{cccc}
    H_1 & 0 & 0 & \cdots \\
    0 & H_2 & 0 &  \\
    0 & 0 & \ddots &  \\
    \vdots & & & H_M
    \end{array}\right )
\end{equation}
where $N\leq M$ and $H_i$ are non-negative and irreducible. Note that only the diagonal blocks are non-zero since $PHP^{-1}$ is Hermitian. Note also that $H_i$ are also Hermitian and thus have a unique largest magnitude eigenvalue that is positive. Each block $H_i$ has a unique positive ground state $|\psi_i\rangle$. Only $N$ of these will correspond to the largest eigenvalue of $\hat{H}$, otherwise the ground state degeneracy would be greater than $N$. Therefore there are $N$ non-negative ground states $P |\psi_i\rangle$ for the Hamiltonian $H$. These states necessarily have distinct support with respect to the computational basis since the basis vectors are non-negative and orthogonal.

\subsection{Permutation matrices}\label{ap: permutations}

In our proof we use the following properties of permutation matrices in order to show that the eigenvalues of the modular matrices form complete sets of roots of unity.

\textbf{Theorem:} Every permutation of a finite set can be written as a product of disjoint cycles.

\textbf{Proof:} Let $\pi$ be the permutation on the finite set $A = \{1,2,\ldots,N\}$. Pick any element $a_1 \in A$. Generate the elements $a_m = \pi^{m-1}(a)$, i.e. $a_2 = \pi(a_1)$, $a_3 = \pi(a_2)$, etc. Then since $A$ is finite the set $\{a_1,\ldots a_M\}$ must be finite with $M\leq N$ and $\pi^M(a_1) = a_1$ and so $(a_1 a_2 \cdots a_M)$ is a cycle. If we have not exhausted the elements of $A$ then we can pick a new element $b_1 \in A$ that is not in $\{a_1,\ldots a_M\}$. Now $\pi^j(b_1) \notin \{a_1,\ldots a_M\}$ for any $j$, since if it were there would exit a $j$ such that $\pi^j(b_1) = a_1$, contradicting that $b_1 \notin \{a_1,\ldots a_M\}$. We therefore generate a disjoint cycle from $b_1$, i.e. $(b_1,\cdots b_K)$. Since $A$ is finite we can repeat this procedure until we have exhausted all elements in $A$.

\textbf{Theorem:} The eigenvalues of the matrix representation of a cycle are a complete set of roots of unity.

\textbf{Proof:} Let $P$ be the matrix representing the cyclic permutation $\pi$ with order $n$. We have that $P^n=\mathds{1}$ and so the matrix has the characteristic polynomial $\lambda^n = 1$. Therefore $P$ has eigenvalues $\{e^{2\pi i \frac{k}{n}} \}^{n-1}_{k=0}$.

\textbf{Corollary:} Every permutation matrix has eigenvalues that form complete sets of roots of unity.

\textbf{Proof:} Since every permutation on a finite set can be decomposed into a product of disjoint cycles, the matrix $P$ can be brought into block diagonal form by a permutation matrix where each block $P_i$ corresponds to a cyclic permutation. Let $m$ be the number of cycles. Each block therefore has eigenvalues of the form $\{e^{2\pi i \frac{k}{n_j}} \}^{n_j-1}_{k=0}$ and so the matrix $P$ has eigenvalues
\begin{equation}
    \{e^{2\pi i \frac{k}{n_1}} \}^{n_1-1}_{k=0} \cup \{e^{2\pi i \frac{k}{n_2}} \}^{n_2-1}_{k=0} \cup \cdots \cup \{e^{2\pi i \frac{k}{n_m}} \}^{n_m-1}_{k=0}.
\end{equation}

\textbf{Theorem:} A non-negative unitary matrix is a permutation matrix.

\textbf{Proof:} Let $U$ be the unitary matrix. Then consider the first non-zero element in the $m^{th}$ column occurring in row $j_m$. Then since all of the columns are non-negative, for this column to be orthogonal to all other columns we must have $U_{j_m,i} = 0$ for $i\neq m$. That is, each row and each column has exactly one non-zero element. Since the matrix is non-negative and unitary this non-zero value is equal to 1. The matrix is therefore a permutation matrix.

\section{TQFT Basics}\label{ap: TQFT}

\begin{figure*}[t!]
    \centering
    \subfigimg[height=.62\columnwidth]{\textbf{(a)}}{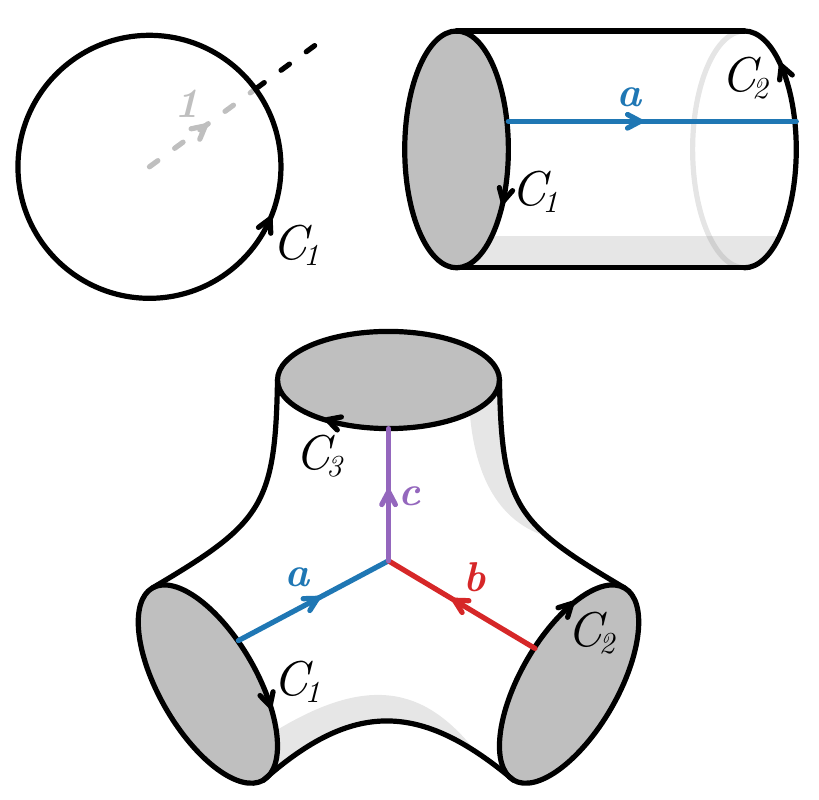}\qquad
    \subfigimg[height=.62\columnwidth]{\textbf{(b)}}{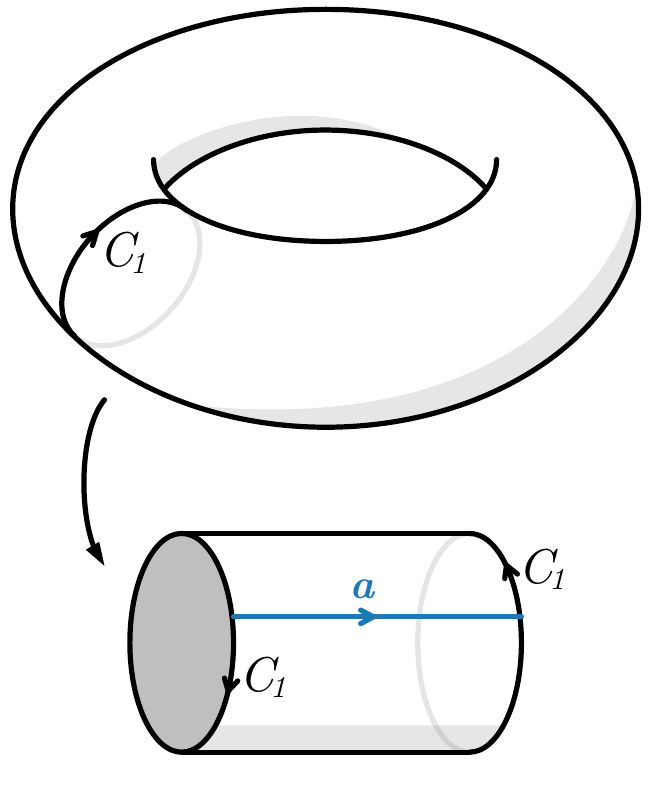}\qquad
    \subfigimg[height=.62\columnwidth]{\textbf{(c)}}{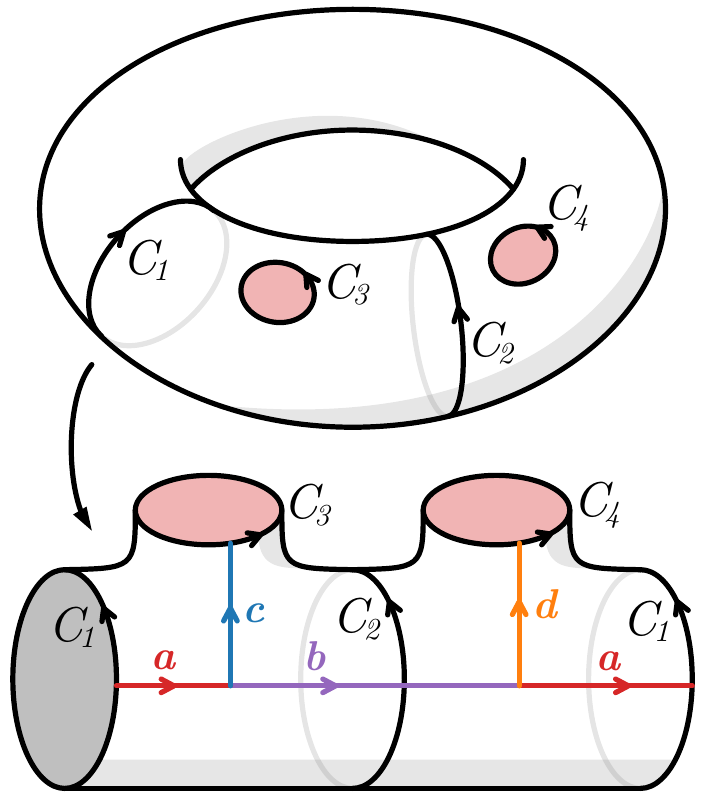}
    \caption{(a) Different elements of the DAP decomposition. Any 2D surface can be decomposed into discs, annuli, and "pants". The labelling along the bounding curves is fusion consistent if it satisfies the constraints in the main text, which are equivalent to considering anyons threading these loops and satisfying the fusion rules of the underlying TQFT. (b) Example DAP-decomposition for the torus. (c) Example DAP-decomposition for the twice-punctured torus. The punctures bind topological fluxes of type $a\times \bar{b}$ and $b \times \bar{a}$ for an abelian theory.}
    \label{fig: DAP}
\end{figure*}

Here we give a very brief review of the basics of TQFT relevant for our work. It will be far from comprehensive and we point the reader to Refs.~\cite{Kitaev,Bonderson2007,Bernevig2015,SteveTopo,Koenig2010,Beverland2016} for more complete discussions. We include this discussion here so that we can more formally state some of the properties used in the main text. We follow most closely the discussion in Ref.~\cite{Beverland2016}.

Let us consider a 2-dimensional orientable surface $\Sigma$, then a $(2+1)$-dimensional TQFT assigns a Hilbert space $\mathcal{H}_\Sigma$ to this surface that depends only on the topology of the surface, i.e., $\mathcal{H}_\Sigma$ does not change with continuous deformations of $\Sigma$. This Hilbert space corresponds to the ground state manifold of our Hamiltonian of interest, but here we do not refer to any Hamiltonian, either on a lattice or in the continuum. A TQFT also assigns a linear map to cobordisms, but we will not explicitly need to refer to cobordisms in our discussion~\cite{Atiyah1988,SteveTopo}. A TQFT is then defined by a set of data that tell us how to label these states in terms of anyon types and the allowed fusion, as well as how these states transform under the mapping class group of the surface. Under the assumption of non-degenerate braiding, this data defines a unitary modular tensor category (UMTC). These anyon labels also correspond to the elementary excitations of the model. 

We proceed by splitting our discussion into three parts: (1) introducing the anyon types, the Verlinde algebra and the Wilson loop operators; (2) labelling states with the DAP ("pants") decomposition; (3) modular transformations, and the mapping class group.

\subsection{Anyons, the Verlinde Algebra, and Wilson Operators}

The starting point is a set of anyon types $\mathbb{A} = \{1,a,b,\ldots\}$. This set contains a unique identity, or vacuum labelled 1, and for each $a$ there is a unique antianyon $\bar{a}$ (it is allowed to have $\bar{a} = a$). We then define the fusion rules for the anyons
\begin{equation}\label{eq: fusion}
    a \times b = \sum_c N^c_{ab} c,
\end{equation}
with $c\in \mathbb{A}$, where $N^c_{ab}$ are non-negative integers called the fusion multiplicities. We restrict to the case $N^c_{ab} \in \{0,1\}$. This defines a fusion category if the fusion is commutative ($a\times b=b\times a$), associative ($(a\times b) \times c) = a \times (b \times c)$), fusion with identity is trivial ($a\times 1 = a$), and that $\bar{a}$ is the unique element that fuses with $a$ to give the identity as one of its fusion channels ($a\times \bar{a} = 1 + \cdots$). The fusion is abelian if there is a unique fusion channel for each pair of anyons and is non-abelian otherwise. The fusion multiplicities also define a Verlinde algebra, spanned by elements $\{w_a\}_{a\in\mathbb{A}}$ such that $w^\dag_a = w_{\bar{a}}$ and the multiplication has the same form as in Eq.~\eqref{eq: fusion}. 

The next ingredient in the definition of a TQFT is the S-matrix. We restrict to the case where the $S$-matrix is unitary (corresponding to non-degenerate braiding) and its matrix elements satisfy $S_{ab} = S_{ba} = S^*_{\bar{a}b}$. The elements of the S-matrix form a representation of the Verlinde algebra and simultaneously diagonlise the matrices with elements $[N_a]_{bc} = N^c_{ab}$. The fusion multiplicities and the S-matrix are related by the Verlinde relation 
\begin{equation}\label{eq: Verlinde relation}
    N^c_{ab} = \sum_{x} \frac{S_{ax} S_{bx} S_{cx}^*}{S_{1x}}.
\end{equation}
The S-matrix also contains the quantum dimension for the anyons: $S_{1a} = d_a/\mathcal{D}$, where $d_1 = 1$, $d_a\geq 1$ and $\mathcal{D}^2 = \sum_a d_a^2$. For abelian theories we have that $d_a = 1$ for all $a \in \mathbb{A}$.

To define the vector space on a surface we introduce the Wilson loop operators for each anyon type. For each directed closed curve $C$ on the surface we have a set of Wilson loop operators $\{\hat{W}_a(C)\}_{a\in\mathbb{A}}$ that form a faithful representation of the Verlinde algebra, i.e., 
\begin{equation}
    \hat{W}_a(C) \hat{W}_b(C) = \sum_c N^c_{ab} \hat{W}_c(C),
\end{equation}
with $\hat{W}_a(C)^\dag = \hat{W}_{\bar{a}}(C) = \hat{W}_a(C^{-1})$. These operators preserve the Hilbert space $\mathcal{H}_\Sigma$, in other words, they preserve the ground state manifold of the parent Hamiltonian. These operators have path-invariance when acting on the ground state manifold, that is, let $C$ and $C'$ be two curves that can be continuously deformed into each other, then $\hat{W}_a(C) |\phi\rangle = \hat{W}_a(C') |\phi\rangle$ for all $|\phi\rangle \in \mathcal{H}_\Sigma$. Also, for any closed loop that is topologically trivial (can be continuously contracted) we have that $\hat{W}_a(C_{trivial})|\phi\rangle = d_a |\phi\rangle$ for all $|\phi\rangle \in \mathcal{H}_\Sigma$. In the main text, these properties of the Wilson loop operators are taken as the defining properties for a lattice Hamiltonian to have a low energy TQFT description.

Next we introduce the Kirby loop projectors along each curve $C$,
\begin{equation}
    \hat{\Omega}_a(C) = S_{1a} \sum_b S_{\bar{a} b} \hat{W}_b(C).
\end{equation}
The set $\{\hat{\Omega}_a(C)\}_{a\in \mathbb{A}}$ is the unique complete set of orthogonal idempotents that span the Verlinde algebra (up to permutations)~\cite{Beverland2016}. These operators have the interpretation of projecting on states with topological flux $a$ threading the curve $C$. For any contractible loop $C$ we have that $\hat{\Omega}_a(C)|\phi\rangle = \delta_{1,a} |\phi\rangle$ for all $|\phi\rangle \in \mathcal{H}_\Sigma$, and so the ground state manifold is such that any region homeomorphic to a disc is flux free.

\subsection{DAP-decomposition}

Equipped with the Wilson loop operators and the Kirby projectors, we are now in a position to label the states in the Hilbert space $\mathcal{H}_\Sigma$ and to define a canonical basis. To do so, we introduce the DAP-decomposition ("pants"-decomposition) of the surface. This consisits of a minimal set of non-intersecting closed curves $\mathcal{C} = \{C_j\}$ that cut the surface into subsurfaces homeomorphic to discs, annuli, and pants (three-punctured spheres), see Fig.~\ref{fig: DAP}(a). 

For a given decomposition we can define a canonical basis, with states labelled by $|a,b,c, \cdots\rangle$, where the anyon labels $a, b, c, \ldots \in \mathbb{A}$ are associated to the curves $C_1, C_2,C_3, \ldots \in \mathcal{C}$, respectively. These states are such that they are eigenstates of the Kirby loop operators of a given type along each curve, e.g., $\hat{\Omega}_x(C_2) |a,b,c,\cdots\rangle = \delta_{b,x} |a,b,c,\cdots\rangle$, and we can write the labelling for a given curve as $l(C_i)$, e.g., $l(C_2) = b$. However, we cannot freely assign any label to each curve in $\mathcal{C}$: the labelling, and so the basis states, must be fusion consistent. A labelling is fusion consistent, if: (a) any disc has trivial labelling, (b) for an annulus between $C_1$ and $C_2$ orietented such that the annulus is to the left of the curves, then $l(C_1) = \widebar{l(C_2)}$, (c) for a pair of pants between $C_1,C_2,C_3$, the labelling is consistent if 
\begin{equation}
    N^{\widebar{l(C_3)}}_{l(C_1) l(C_2)} \neq 0.
\end{equation}
This labelling of a curve $l(C)$ can equivalently be viewed as an anyon of type $l$ threading the curve $C$. We can keep track of the labelling by drawing world lines on the surface that obey the anyon fusion, see Fig.~\ref{fig: DAP}. A given fusion diagram on the surface corresponds to an element of the vector space $\mathcal{H}_\Sigma$. The dimension of the Hilbert space $\mathcal{H}_\Sigma$ (ground state degeneracy) for the torus is then given by the number of anyon types, for the three-punctured sphere it is $\sum_{a,b,c}N^c_{ab}$, and similarly for other surfaces.

\begin{figure}[b!]
    \centering
    \includegraphics[width=\columnwidth]{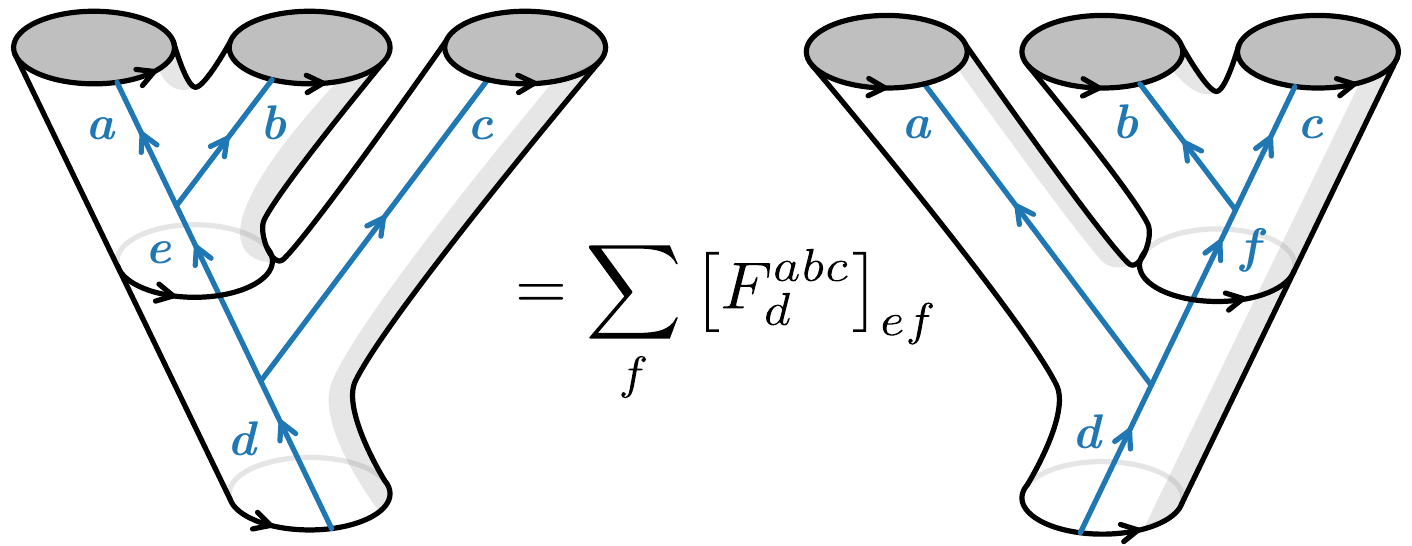}
    \caption{The F-moves relate two different labellings of the four-punctured sphere, corresponding to the DAP-decompositions $\mathcal{C}$ and $\mathcal{C}'$. The $F$-moves define associativity of fusion on the level of the ground state Hilbert space.}
    \label{fig: modular F}
\end{figure}

\begin{figure}[t!]
    \centering
    \includegraphics[width=\columnwidth]{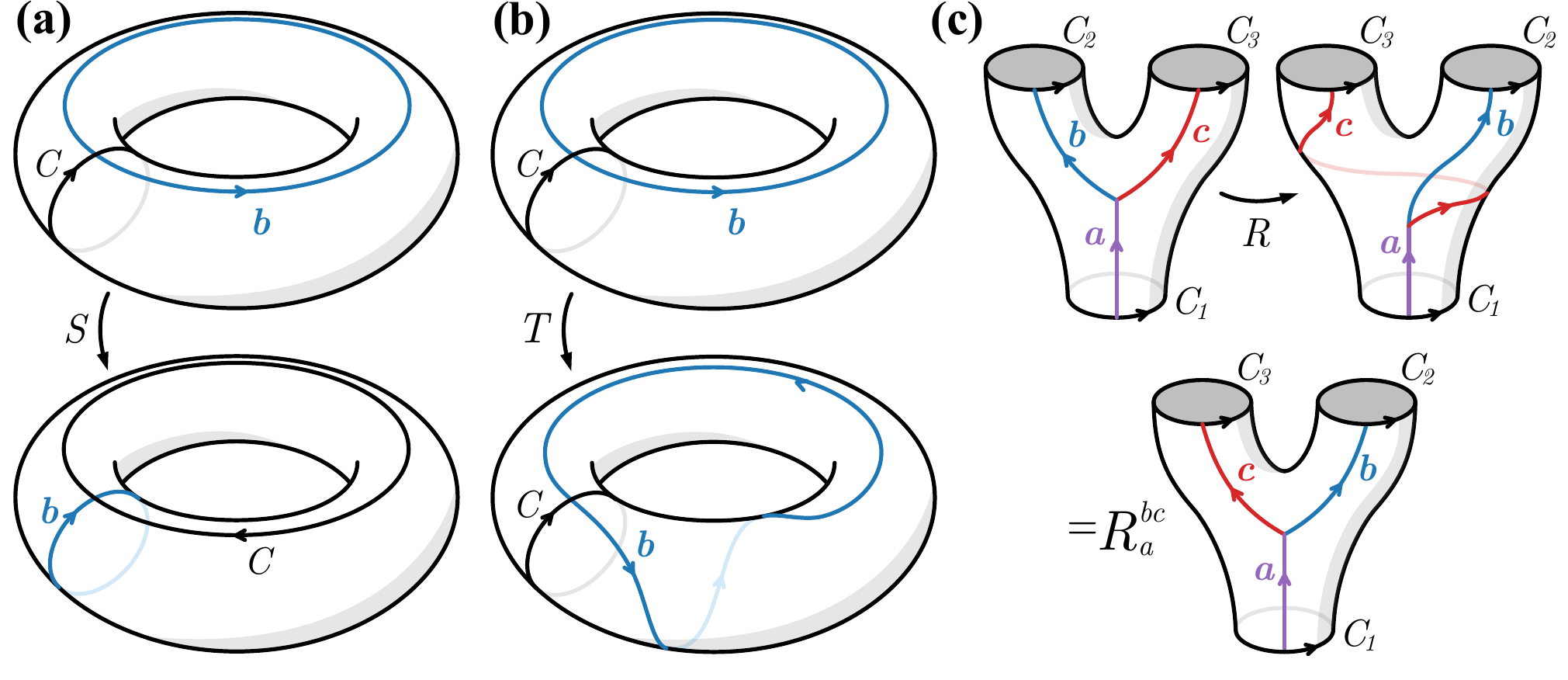}
    \caption{Modular transformations that take a surface back to itself. (a) The $S$-matrix, which swaps the meridian and longitude of the torus. (b) The Dehn twist corresponding to cutting the torus, performing a full twist and gluing it back together again. (c) The $R$-matrix corresponding to swapping two of the punctures in the three-punctured sphere.}
    \label{fig: modular S T R}
\end{figure}

Now that we have a basis for our vector space we need to be able to relate different DAP-decompositions. This is done with the $S$-matrix and $F$-moves. The $S$-matrix relates two inequivalent DAP-decompositions for the torus, see Fig.~\ref{fig: modular S T R}(a). We can equally well choose to label states along the meridian of the torus with DAP-decomposition $\mathcal{C}_{m}$ or along the longitude $\mathcal{C}_{l}$. The corresponding basis states are related by 
\begin{equation}
    |a\rangle_{l} = \sum_b S_{ab} |b\rangle_{m},
\end{equation}
or equivalently, $\langle a |_{l} |b\rangle_{m} = S_{ab}$.
We can also choose two different decompositions for the four-punctured sphere, as shown in Fig.~\ref{fig: modular F}. These are related by the so-called $F$-moves, which are the matrix elements $[F^{abc}_d]_{ef}$. These matrices are unitary and define the associativity for the direct product of the vector spaces of pants diagrams (gluing two pants diagrams to make a four punctured sphere). These $F$-moves are part of the data of the TQFT and must satisfy the pentagon equation, which is the consistency equation for the associativity of fusion.

At this point we can show that the property for moving Wilson loops across flux free regions follows from the fusion consistent labelling of the states in the TQFT. In particular, consider a pair-of-pants with one of the punctures with a trivial labelling ($\hat{\Omega}_1(C_1)|\phi\rangle = |\phi\rangle$ for $C_1$ surrounding the puncture). The labelling either side of the puncture ($C_2, C_3$) must then be the same. This in turn means that the action of the Wilson loop operators along $C_1$ and $C_2$ must be the same, i.e., $\hat{W}_b(C_2)|1,a,a\rangle = \hat{W}_b(C_3) |1,a,a\rangle$ for all $a,b \in \mathbb{A}$.

\subsection{Modular Transformations and the Mapping Class Group}

Now that we have defined how vector spaces are assigned to surfaces, the remaining ingredients in the definition of a TQFT determine how these states transform if we perform deformations to the surface that bring it back to its original configuration. For the torus these are the modular transformations and for more general surfaces $\Sigma$, these are the elements of the mapping class group $MPG_\Sigma$ for that surface.

For the torus $T^2$, the mapping class group is the group of modular transformations, i.e., $MPG_{T^2} \cong SL(2,\mathbb{Z})$. This group is generated by two-elements $s$ and $t$. The first corresponds to the $S$-matrix and swaps the meridian and longitude of the torus, which we met earlier. The second corresponds to the $T$-matrix and performs a Dehn twist on the torus. The Dehn twist is performed by cutting the torus along a curve (let's say the meridian) and twisting by one full rotation then gluing back together, arriving back at the torus. Let us consider labelling of the states on the torus with respect to the meridian (along another curve), then the states before and after will have the same labelling, see Fig.~\ref{fig: modular S T R}(b). However, there is freedom that this process could have introduced a state dependent phase, i.e. $\hat{T}|a\rangle_m = e^{-2\pi i c_-/24} \theta_a|a\rangle_m$. The phase $\theta_a = e^{2\pi i h_a}$ is known as the topological spin (and $h_a$ the conformal scaling dimension) and $c_-$ is chiral central charge. The chiral central charge is related to the topological spins via $e^{2\pi i c_-/8} = \mathcal{D}^{-1}\sum_a d_a^2 \theta_a$, which specifies the $c_-$ mod 8~\cite{Kitaev2006}. The $T$-matrix is the diagonal matrix $T_{ab} = e^{-2\pi i c_-/24}\theta_a \delta_{ab}$, and is part of the definition of the TQFT. The MPG for the torus is generated by these two elements. It is also believed that the $S$ and $T$ matrix fully specify a TQFT. As we have defined them, the $S$ and $T$ matrix generate a projective representation of the modular group and satisfy $(ST)^3 = C$, and  $S^2 = C$, where $C$ is the conjugation operator corresponding to flipping both spatial directions. It is also common to define the $T$-matrix without the chiral central charge, i.e., $T_{ab} = \theta_a \delta_{ab}$, which has the result that $(ST)^3 = e^{2\pi i c_-/8}C$. For the majority of the main text we consider non-chiral theories where $c_- = 0$ mod 24, but consider chiral theories in Sec.~\ref{sec: chrial}.

As well as the torus, the mapping class group for the pair-of-pants surface $\Sigma$ is an important ingredient of the TQFT. The labelling of states on this surface are shown in Fig.~\ref{fig: modular S T R}(c). Here we can also move one of the punctures and swap it with one of the others. In this way we return to the same surface but the labelling has changed. This operation corresponds to braiding as specified by the elements $R_{a}^{bc}$. More precisely, if we denote the operation, $\hat{R}_{2,3}$ as the one that swaps curves $C_2$ and $C_3$, then we have $\hat{R}_{2,3}|a,b,c\rangle = R_a^{bc} |a,c,b\rangle$. These $R$-moves must be consistent with the associativity of fusion and satisfy the hexagon equations.
For more general surfaces with higher genus and more punctures, the mapping class group can be generated by the S-matrix, Dehn-twists, and braids. A useful operation in this context is the braid operator $\hat{B} = \hat{F}^{-1} \hat{R} \hat{F}$, which braids a pair of punctures on an M-punctured surface, see Ref.~\cite{Beverland2016} for more details.

It is important to note that here we simply list the data that defines a TQFT, namely given by the anyon types $\mathbb{A}$, the fusion multiplicities $N^c_{ab}$, the $S$, $T$ matrices and the $F$ and $R$ moves. This data is, however, not all independent. We saw an example of this in Eq.~\eqref{eq: Verlinde relation} which says that the fusion multiplicities can be derived from the $S$-matrix. Please see e.g., Refs.~\cite{SteveTopo,Kitaev2006,Bonderson2007,Bernevig2015} for more details of how these data are related and what conditions must be satisfied for them to be consistent and define a TQFT.

\section{Generating Canonical Basis States}\label{ap: orthogonality}

As well as the implicit definition of the basis states as eigenstates of the different Kirby projectors, it will be useful to have a more constructive definition of these basis states. This will be used to show properties of our construction for the transformation $\hat{U}$, which we do for abelian models in the main text. For this we use a property of the Wilson loop operators that holds for abelian TQFTS, namely that for two non-contractible loops, $C$ and $C'$ that cross exactly once we have that
\begin{equation}\label{eq: crossing Wilson}
    \hat{W}_a(C) \hat{W}_b(C') |\phi\rangle = \mathcal{D} S_{ab} \hat{W}_b(C')\hat{W}_a(C) |\phi\rangle,
\end{equation}
for all $|\phi\rangle \in \mathcal{H}_\Sigma$. With this we can show that the vertical (meridian) canonical basis states on the torus can be generated as follows
\begin{equation}
    |a\rangle_m = \hat{W}^h_a |1\rangle_m,
\end{equation}
where $\hat{W}^h_a$ is any horizontal Wilson loop and $W^v_a|1\rangle_m =  |1\rangle_m$ for all $a \in \mathbb{A}$. This matches the canonical DAP labelling since
\begin{equation}
\begin{aligned}
     \hat{\Omega}_{a}^v |b\rangle_m &=  \frac{1}{\mathcal{D}}\sum_c S_{ac}^* \hat{W}_c^v \hat{W}_b^h |\mathds{1}\rangle_m \\
    &= \sum_c S_{ac}^* S^{}_{cb} \hat{W}_b^h \hat{W}_c^v |\mathds{1}\rangle_m \\
    & = \sum_c S_{ac}^* S^{}_{cb}\hat{W}_b^h  |\mathds{1} \rangle_m \\
    & = \delta_{ab} |b\rangle_m,
\end{aligned}
\end{equation}
and so these states are eigenstates of the Kirby projector with the corresponding label and annihilated by all other Kirby projectors.

Note that the above also applies to the case with punctures considered in Fig.~\ref{fig: flux binding}. This procedure generates orthogonal states in this case as well since it relies only on the property in Eq.~\eqref{eq: crossing Wilson}, which still holds away from the punctures. Given a trivial state $|1\rangle$, these states form an orthonormal basis. However, we need the additional arguments or modified procedure presented in the main text to argue that the relative phases of the states generated using curves $C_1$ and $C_2$ are the same. In general this need not be the case and there can be a state dependent phase between those generated using $C_1$ versus $C_2$. However, if the region between these two curves has trivial topological flux, then the Wilson loop operators along $C_1$ and $C_2$ must have the same action up to some global phase.

\section{Proof: $\mathbb{Z}_N$ string-net sign-problems}

Here we prove that all $\mathbb{Z}_N$ string-net models, except generalized toric code models, have intrinsic sign problems. For abelian string-nets built on $\mathbb{Z}_N$ ($N\geq 2$) we have $N$ distinct theories labelled by $p = 0,\ldots, N-1$, and we refer to the corresponding string-net model as $\mathbb{Z}_N^p$. These models have $N^2$ anyon excitations labelled by the pairs $(s,m)$ with $s,m=0, \ldots, N-1$. As shown in Ref.~\cite{Lin2014}, the topological spins in the $T$-matrix have the form
\begin{equation}
    \theta_{(s,m)} = \exp\left\{ i 2 \pi \left(\frac{ps^2}{N^2} + \frac{ms}{N}\right) \right\}.
\end{equation}
For examples, for $\mathbb{Z}_2$ there are two theories, $\mathbb{Z}^0_2$ and $\mathbb{Z}_2^1$, which are the toric code and double semion model, respectively. The first has topological spins $\theta = 1,1,1,-1$ and the second $\theta = 1, i ,-i, 1$.

We will now prove that for all $N$ and $p \neq 0$, $\mathbb{Z}_N^p$ has an intrinsic sign problem by showing that the topological spins do not form complete sets of roots of unity. Note that for $p=0$ the model is trivially stoquastic in the standard qudit basis and does not have a sign problem and corresponds to a $\mathbb{Z}_N$ generalization of the toric code. Correspondingly we have that $\theta_{(s,m)} = e^{i\frac{2\pi}{N}ms}$, which do form complete sets of roots of unity. We split the remaining values of $p$ into those for which $p$ and $N$ are coprime and those that are not.

We start with the case where $p$ and $N$ are coprime, i.e., the greatest common divisor $GCD(N,p) = 1$. We first note that when $s=0$ we have $\theta_{(0,m)} = 1$ for all $m$, i.e. $N$ topological spins are $+1$. We also note that $\theta_{(s=1,m=0)} = e^{i \frac{2\pi}{N^2}p}$. However, since $p$ and $N$ are coprime we have to take the $N^2$ power to get back to unity, i.e., $\theta_{(1,0)}^{n} \neq 1$ for all $n<N^2$. This means one of the sets of roots of unity would have to contain $N^2$ elements, but this is a contradiction since there are only $N^2$ topological spins and we already know $N$ of them are equal to 1. 

For the case when $p$ and $N$ are not coprime let us define $\tilde{N} = GCD(p,N)>1$ and $p = \alpha \tilde{N}$, $N = \beta \tilde{N}$. In this case we have $\theta_{(1,0)} = e^{i \frac{2\pi}{N^2}p} = e^{i \frac{2\pi}{N}\frac{\alpha}{\beta}}$ and that $\theta_{(1,0)}^\beta = e^{i 2 \pi \frac{\alpha}{N}}$. Now since $\alpha$ and $N$ are coprime we must have that $\theta_{(s,m)} = e^{i\frac{2\pi}{N}}$ for some values of $s$ and $m$. We will now show that this is not the case. Looking again at the general form we can write it as
\begin{equation}
    \theta_{(s,m)} = \exp\left\{ i \frac{2 \pi}{N} \left(\frac{\alpha s^2}{\beta} + ms\right) \right\} = \exp\left\{ i \frac{2 \pi}{N} f_p(s,m) \right\},
\end{equation}
so we now need to find $s$ and $m$ such that $f_p(s,m) = 1 \text{ mod } N$. For $f_p(s,m)$ to be an integer we first need that $s = k \beta$ for some $k=0,\ldots,\tilde{N}-1$. Then we can write the integer values as
\begin{equation}
    f_p(s,m) = \beta k (k + m)
\end{equation}
However, both $f_p(s,m)$ and $N$ are divisible by $\beta$ for all values of $k$ and $m$. Therefore, there are no such values of $s$ and $m$ and we have an incomplete set of roots of unity. In conclusion, \emph{all $\mathbb{Z}_N^p$ string-net models with $p\neq0$ have an intrinsic sign problem}.

\section{Topological flux bound to lattice dislocations}\label{ap: flux binding}

In Sec.~\ref{sec: LSAP} we considered the possibility that the lattice dislocations we introduce could bind a topological flux. In the main text we were able to account for the case where a dislocation binds a unique topological flux by a simple modification of the procedure, as illustrated in Fig.~\ref{fig: flux binding}. In this section we also account for the possibility that the dislocation doesn't bind a unique flux but a superposition of fluxes, and show that this case can be removed by a local stoquastic perturbation. We allow the freedom for such a local stoquastic perturbation in our procedure for $\hat{U}$ in the main text, see Sec.~\ref{sec: assumptions}.

Let us consider the possibility that a defect binds a superposition of two anyon types and for simplicity stay in the case where the Wilson operators have finite width support and are exactly commuting. In this case the ground state in the presence of a defect and an anti-defect looks like $|\psi\rangle = \alpha|a\rangle + \beta|b\rangle$, where $|a\rangle$ corresponds to a state with a unique topological flux of type $a$ bound to one of the dislocations. The state $|a\rangle$ is such that for a curve $C$ enclosing this dislocation but not the other, we have $\hat{\Omega}_c(C)|a\rangle = \delta_{ac}|a\rangle$, and similarly for $|b\rangle$. Since the Wilson loops along $C$ commute with the Hamiltonian, so does the Kirby loop operator, meaning that there is no matrix element in the Hamiltonian between $|a\rangle$ and $|b\rangle$. Therefore, these two states are separately ground states of the Hamiltonian and are exactly degenerate. However, these two states can be differentiated by a local operator, namely, the Kirby projectors on the curve $C$ surrounding the dislocation. We can therefore apply a local perturbation near the defects to lift the degeneracy and chose a single anyon type. 

The remaining loose end to tie is whether this degeneracy cannot be lifted by any stoquastic perturbation. Let us assume that no local stoquastic operator lifts this degeneracy. We have, however, that the Kirby loops operator ($\Omega_a(C)$), associated with one of the fluxes allowed in the degeneracy (say $|a\rangle$, without loss of generality), can lift this degeneracy (by adding it with a minus sign to the Hamiltonian). We argue that this implies that only a purely imaginary operator can lift the degeneracy. Let us split this Kirby operator into a sum of three operators: A stoquastic part $O_s$ which includes all the diagonal entries and all the negative off diagonal ones, a part $O_{as}$ including all the positive off diagonal entries, and a part $O_i$ including all the imaginary entries. Since we assumed that no stoquastic operator can split the degeneracy $O_s$ can be removed from $\Omega(a)$ without affecting the splitting---indeed all its eigenvalues must be equal in this degenerate subspace and hence it must act at the identity. Similarly since $-O_{as}$ is stoquastic, $O_{as}$ can be removed as well. This leaves the task of splitting the degeneracy solely on $O_i$, the imaginary part of that Kirby operator on the computational basis.  

This remaining scenario can be seen a form of spontaneous breaking of complex conjugation symmetry in statistical mechanics: A situation where the obvious/trivial complex conjugation symmetry of statistical mechanics becomes spontaneously broken such that an infinitesimal complex conjugation breaking perturbation (as $O_i$ above) picks up a unique ground state which strongly breaks complex conjugation symmetry while all other complex conjugation symmetry respecting perturbation leave this degeneracy intact. Focusing on doubly degenerate states, it was shown in Ref.~\cite{Ringel2017} that such a scenario is impossible. In a nut shell, they considered the off-diagonal expectation of $O_i$ on the ergodic basis implied by the degeneracy. It was shown that the off-diagonal element of the operator $|O_i|$ (defined as the element-wise absolute value of $O_i$) bounds the former whereas, on the other hand, it must be zero other wise $-|O_i|$, a stoquastic operator, can lift the degeneracy. We leave extensions to 3-fold or higher degeneracies to future work.

\end{document}